\documentclass[a4paper,12pt]{article}
\usepackage[top=1.00in, bottom=1.00in, left=1.00in, right=1.00in]{geometry}

\usepackage[english]{babel}
\usepackage[utf8]{inputenc}
\usepackage[T1]{fontenc}
\usepackage{natbib}

\usepackage[section]{placeins}
\usepackage{setspace}

\usepackage{enumitem}
\setlist{noitemsep}  

\usepackage{amscd,amsmath,amssymb,amsfonts,amsthm,bbm,latexsym,mathrsfs,mathtools}
\usepackage[subnum]{cases}	
\usepackage{arydshln}   
\usepackage{dsfont}

\def\R {\mathds{R}}

\def\bc {\mathbf{c}}

\def\by {\mathbf{y}}

\def\balpha {\boldsymbol{\alpha}}
\def\bbeta  {\boldsymbol{\beta}}
\def\bgamma {\boldsymbol{\gamma}}
\def\bdelta {\boldsymbol{\delta}}
\def\bepsilon {\boldsymbol{\epsilon}}
\def\bzeta {\boldsymbol{\zeta}}
\def\btheta {\boldsymbol{\theta}}
\def\biota {\boldsymbol{\iota}}
\def\bmu    {\boldsymbol{\mu}}
\def\brho   {\boldsymbol{\rho}}
\def\bsigma {\boldsymbol{\sigma}}
\def\btau {\boldsymbol{\tau}}

\makeatletter
\newcommand*{\distas}[1]{\mathbin{\overset{#1}{\kern\z@\sim}}}	
\makeatother
\newcommand*\abs[1]{\left|#1\right|}		

\theoremstyle{definition}
\newtheorem{assumptions}{Assumption}
\theoremstyle{remark}
\newtheorem{remark}{Remark}
\theoremstyle{plain}

\RequirePackage[colorlinks=true, citecolor=blue, urlcolor=blue]{hyperref}

\usepackage{xcolor,subfigure,epsfig,epstopdf,rotating,graphics,graphicx} 
\graphicspath{{./figures/}}
\usepackage[width=0.90\linewidth, font={footnotesize}]{caption}

\usepackage{rotating,lscape}
\usepackage{booktabs,longtable,float,array,multirow,colortbl,adjustbox}
\usepackage{pdflscape}

\newcolumntype{C}[1]{>{\centering\arraybackslash}p{#1}}

\usepackage[colorinlistoftodos,color=orange!60]{todonotes}
\usepackage{regexpatch}
\makeatletter
\xpatchcmd{\@todo}{\setkeys{todonotes}{#1}}{\setkeys{todonotes}{inline,#1}}{}{}
\makeatother

\newlist{todolist}{itemize}{2}
\setlist[todolist]{label=$\square$}
\usepackage{pifont}
\usetikzlibrary{shapes,arrows}	
\usepackage{tkz-berge}
\usetikzlibrary{positioning}
\usepackage{pgfplots} 
\usepackage{pgfplotstable}
\usetikzlibrary{positioning,calc}
\usetikzlibrary{backgrounds}
\usetikzlibrary{shapes, arrows}
\usetikzlibrary{positioning}
\usetikzlibrary{matrix}
\usetikzlibrary{plotmarks}
\usetikzlibrary{arrows.meta}
\usepgfplotslibrary{groupplots}
\pgfplotsset{compat=newest}
\usepgfplotslibrary{dateplot}

\title{\vspace{-60pt} \textbf{Bayesian SAR model with stochastic volatility and multiple time-varying weights}
}

\author{
Michele Costola\thanks{Ca' Foscari University of Venice, Italy. (corresponding author) \color{blue}\texttt{michele.costola@unive.it}}
\and
Matteo Iacopini\thanks{Queen Mary University of London, United Kingdom.   \color{blue}\texttt{m.iacopini@qmul.ac.uk}} 
\and
Casper Wichers\thanks{ADC, Data \& AI Consultancy, The Netherlands. \color{blue}\texttt{casperwichers@gmail.com}}
}

\date{\today}

\begin{document}

\maketitle

\begin{abstract}
A novel spatial autoregressive model for panel data is introduced, which incorporates multilayer networks and accounts for time-varying relationships.
Moreover, the proposed approach allows the structural variance to evolve smoothly over time and enables the analysis of shock propagation in terms of time-varying spillover effects.

The framework is applied to analyse the dynamics of international relationships among the G7 economies and their impact on stock market returns and volatilities.
The findings underscore the substantial impact of cooperative interactions and highlight discernible disparities in network exposure across G7 nations, along with nuanced patterns in direct and indirect spillover effects.


\end{abstract}


\doublespacing

\newpage

\section{Introduction}   \label{sec:introduction}

Recent technological advancements, particularly in Natural Language Processing (NLP), have significantly expanded the pool of analysable data by extracting key metrics from news providers and social media.
This development has allowed explaining potential dynamics that might substantially impact the economic and financial landscape.
For example, \cite{baker2016} introduced an Economic Policy Uncertainty (EPU) index, derived from the frequency of newspaper coverage and tends to spike during major political and economic events. 
This index is associated with heightened stock market volatility and reduced investment and employment in policy-sensitive sectors.
Likewise, \cite{caldara2022} have introduced a metric for Geopolitical Risk (GPR), demonstrating its association with fluctuations in the business cycle and GDP contractions. 
Furthermore, additional research has explored diverse subjects, encompassing climate change \citep{engle2020}, pandemics \citep{baker2020}, and energy commodities \citep{abiad2023}.
It is important to highlight that a significant portion of the existing literature primarily focuses on either a single-country perspective or addresses issues on a global scale without distinguishing the specific actors involved.

This article adopts a different approach by examining cross-country relationships and categorising them according to their cooperative or conflictual attributes.
To exemplify, envision a scenario where one country imposes trade barriers on another within a specific industry. The adverse economic repercussions of such a negative event could extend beyond the immediate parties involved, impacting their respective trading partners.

A practical approach to model such relationships is through a network framework, which entails encoding the structure of a directed, weighted network without self-loops.
Network modelling has gained substantial popularity in financial contagion and systemic risks after the global financial crisis \citep{billio2012econometric,diebold2014network, bonaccolto2019estimation,jackson2021systemic}.
Our framework encapsulates the network representation by a square $n\times n$ matrix $W$, where $n$ indicates the number of countries. 
The weights in $W$ (i.e., its nonzero entries) describe the strength of a linkage, which is directed due to the possible absence of reciprocity: country $i$ may undertake an action against country $j$ without a corresponding action from country $j$.
Moreover, the inclusion of diverse relations (either positive or negative) among the same countries may have an asymmetric impact. To accommodate this possibility, we allow the network to be regarded as multilayered, where each layer is an $n\times n$ matrix encoding a specific type of relationship among the same set of countries.

The existing literature in spatial statistics examines models incorporating the interdependencies among observables, represented by one or more spatial weight matrices and potentially cross-sectionally varying strengths of network effects.
In this context, the spatial autoregressive model \citep[SAR, see][]{ord1975estimation,anselin1988spatial} emerges as a natural choice.
However, spatial statistics and econometrics conventionally employ static geographic networks, which implies that the contemporary relationships among observables are assumed to remain constant over time. This contrasts with the dynamic nature of country relationship data.
Moreover, cooperative and conflictual relationships allow the delineation of various types of interconnections, which may exert differing impacts on each country and, consequently, its exposure to the network.

Recently, \cite{debarsy2022bayesian} designed an extension of the SAR model to take multiple layers into account and followed a Bayesian approach to inference. However, this framework has limitations of concern for the motivating application of this article.
First and foremost, the time dimension is neglected as the spatial networks are treated as constant. Second, it is important to note that the innovation variance coefficient is assumed to be identical within the cross-section and constant over time in contrast to the heteroscedasticity commonly present in financial data.
Considering these features enables quantifying the influence of these relationships on the financial and economic system.

In response to the identified desiderata and drawing on recent advances in spatial statistics and financial econometrics, we introduce a novel spatial autoregressive model tailored for panel data \citep[e.g., see][]{anselin2008spatial}.
The proposed model concurrently addresses (i) the incorporation of various types of networks within a multilayer structure, (ii) the consideration of time-varying networks, (iii) the inclusion of country-specific network exposure weights, and (iv) allows for the temporal variation of the structural variance. Additionally, the spatial multiplier matrix enables the analysis of the impact of shocks in terms of both direct and indirect spillover effects. 
Expanding upon the insights provided by \cite{corrado2012economics}, who highlighted the limitations of using weight matrices solely based on spatial location, we develop a model that leverages various forms of cross-sectional dependence.
In the literature, there are two primary approaches when considering multiple-weight matrices. 
One involves using a single matrix per period and applying different criteria for selection or model combination \citep[e.g., see][]{kelejian2016extension,zhang2018spatial}. 
The other strand employs a convex combination of multiple weight matrices to capture the informational content of various dependence structures \citep[e.g., see][]{lee2010efficient,debarsy2018flexible,debarsy2022bayesian}.
This study adopts the latter approach and formulates a spatial autoregressive model with a convex combination of multiple networks to characterise the interdependence among observables. Moreover, the combination weights are inferred from the data, enabling us to discern the relative significance of each network retrospectively.

The approach outlined here is also connected to the work of \cite{bonaccolto2019estimation} and \cite{billio2023impact}, which introduce a static SAR model incorporating multiple but time-constant spatial networks in studying stock returns.
Specifically, \cite{billio2023impact} introduces a single-layer network-augmented linear factor model for asset pricing, whereas \cite{bonaccolto2019estimation} uses multiple network layers to encode different types of pairwise causal relationships estimated in a preliminary first stage.
Conversely, our approach is designed for multiple and time-varying networks and proposes a Bayesian inferential procedure that quantifies uncertainty about all quantities of interest. Moreover, our empirical application is on a macroeconomic scale as it involves multiple observed and weighted networks encoding the type and strength of political relationships among a set of countries.

The use of SAR models in economics and finance has spread significantly over the last decade. For instance, this framework has been successfully applied to define spatial arbitrage pricing theory \citep{hu2023arbitrage} and to model the cross-sectional correlation of abnormal returns \citep{ahlgren2017tests}.
Recently, \cite{yang2021estimation} investigated spatial cointegration through a dynamic panel spatial vector autoregressive model, extending the seminal work by \cite{yu2012estimation}.

In the empirical analysis, we focus on the Group of Seven (G7) economies and examine how the dynamics of their network relationships over time impact their respective stock markets in returns and volatilities.
The relationships within the G7 can be influenced by various factors such as economic policies, trade agreements, and geopolitical events. In times characterised by cooperation or mutual interests, favourable associations have the potential to enhance investor confidence and positively influence stock market indices. Conversely, instances involving trade disputes resulting in trade barriers, heightened political tensions, or economic crises may engender adversarial relations, leading to increased market volatility and the potential for a downturn in stock market performance.
Incorporating positive and negative interactions within a multilayer network can provide valuable insights into how these relationships affect financial markets.
The data on relationships between G7 countries is sourced from the Integrated Crisis Early Warning System (ICEWS) database. The ICEWS data encompasses encoded interactions among socio-political entities involving cooperative or adversarial actions across individuals, groups, sectors, and countries, retrieved through articles and news from various outlets and media platforms. To the best of our knowledge, this is a novel utilisation of the dataset within the financial econometrics and financial network literature.

Our findings reveal an interesting perspective on country relationship data.
First, the layer-specific weights indicate that cooperative interactions substantially influence the stock markets more than conflictual interactions.
This emphasises the cooperative nature of relationships within the G7. 
Second, country-specific spatial weights highlight varying levels of exposure to the network. 
While the United States holds the top position in network centrality, its stock market exhibits a non-significant (though still significant but low) exposure to the network in terms of returns (volatilities), setting it apart from other G7 countries.
This highlights that relationships originating in the United States significantly impact other G7 stock markets, whereas the reverse influence is not observed.
Canada, Japan, and the United Kingdom display substantial coefficients, indicating significant exposure of their financial systems to the network. 
France, Germany, and Italy exhibit even higher levels of network exposure, possibly ascribed to extensive economic interdependencies within the European Union.
The examined spillover interdependencies within the G7 nations underscore the complex interplay between direct and indirect spillover effects, frequently displaying opposite dependence.  
We show that the United States exhibits a negative correlation between spillovers on volatilities and the EPU index, indicating diminished G7 influence on the country during periods of high economic uncertainty.
%

The remainder of this article is organised as follows. 
Section~\ref{sec:model} outlines the proposed models. 
Section~\ref{sec:inference} presents a Bayesian approach to inference. 
Section~\ref{sec:application} provides the results of the empirical application on the G7 group. 
Finally, Section~\ref{sec:conclusion} offers the concluding remarks.

\section{Model}   \label{sec:model}

In the standard multifactor model, the common factors may not be able to link to the interrelationships and the source of heterogeneity in the data.
A possible way of dealing with this issue relies on introducing cross-dependence between the assets as a network combining systematic and idiosyncratic risk components.

Let $\by_t$ denote a $n$-dimensional vector of response variables (e.g., stock indices) and $\mathbf{f}_t$ be a $k$-dimensional vector of covariates (or common factors).
To capture contemporaneous dependence among the response variables, commonly used spatial statistics and econometrics models introduce an exogenous \textit{spatial} matrix $W \in \R^{n\times n}$ that represents the geographical distance between the response variables. More generally, $W$ can describe a network structure that encodes the (contemporary) interconnections among the response variables, such that $W_{i,j} \neq 0$ whenever the $i$th and $j$th responses are connected.
Besides the connectivity structure in $W$, the strength of such spatial relationships is given by a scalar parameter $\rho$, leading to the definition of the relational matrix $A = (I_n - \rho W)$. 
Combining these elements results in the spatial autoregressive (SAR) model \citep[e.g., see][]{anselin1988spatial}:
\begin{equation}
\label{eq:model_SAR}
A \by_t = \balpha_0 + B \mathbf{f}_t + \bepsilon_t, \qquad \bepsilon_t \sim \mathcal{N}_n(\mathbf{0}, \Sigma),
\end{equation}
where $\Sigma = \operatorname{diag}(\sigma_1^2,\ldots,\sigma_n^2)$ is a $n\times n$ positive definite diagonal matrix. Notice that, in the special case where $W=\mathbf{0}_{n}$, a square matrix of zeros, model \eqref{eq:model_SAR} simplifies to a standard linear model.

This approach poses several undesirable limitations. First, it ignores the additional information coming from multiple sources of dependence, which results in several network matrices. Second, assuming a unique spatial weight parameter is particularly restrictive when the network affects the response variables differently. Third, in the presence of panel data, this static model does not account for any time variation of the dependence structure. For instance, real-world financial and social networks typically evolve due to structural breaks or smooth changes \citep{billio2012econometric}.

We propose an extension to the definition of the relational matrix $A$ to address these three major issues. Following \cite{bonaccolto2019estimation}, we introduce multiple sets of networks or \textit{layers}, $W_1,\ldots,W_d$. Using a multilayer network allows us to simultaneously account for different types of dependence (each one encoded by a single network). Moreover, by assigning a weight $\delta_i$ to each layer $W_i$, the ensuing convex combination permits us to infer the relative importance of each type of connectivity in the global cross-sectional dependence scheme.
Motivated by our empirical application to a panel of countries, we replace the single weight parameter with country-specific network weights, $\brho = (\rho_1,\ldots,\rho_n)'$. Then, following \cite{elhorst2003specification}, we define the diagonal matrix $R = \operatorname{diag}(\rho_1,\ldots,\rho_n)$, where the $j$th diagonal element represents the strength of the exposure of country $j$ to the network. This allows us to uncover the cross-sectional heterogeneity of network exposition. We remark that country-specific network exposures $\rho_j$ are of great importance, as they link the political network structure to the financial response variable, providing information on the strength of the dependence between these two variables.
Finally, we take into account the change over time of the dependence structure by incorporating a time series observation for each layer $i=1,\ldots,d$, thus resulting in a time series of multilayer networks, $\{ W_{1,t},\ldots, W_{d,t} \}_{t=1}^T$.

Therefore, we generalise the definition of the relational matrix to:
\begin{equation}
A_t = I_n - R \left(\sum_{i=1}^d \delta_i W_{i,t} \right),
\label{eq:At_def}
\end{equation}
with $R = \operatorname{diag}(\rho_1,\ldots,\rho_n)$. All the entries of the matrices encoding the network structures are non-negative (i.e., $W_{i,t, jk} \geq 0$ for every $i,t$ and every $j \neq k$) and self-loops are excluded ($W_{i,t,jj} = 0$ for every $i,t,j$). Finally, we impose row-normalisation to $W_{i,t}$, for each $i=1,\ldots,d$ and $t=1,\ldots,T$.

Finally, we extend the baseline model by incorporating stochastic volatility, allowing the structural variance to vary with time, as motivated by the observed heteroscedasticity in financial data \citep[e.g., see][]{taylor1994modeling}.
Combining all the elements, the structural form of the model is specified as follows:
\begin{align}
\label{eq:model_structural}
\notag
A_t \by_t & = X_t \bbeta + \bepsilon_t, \qquad \bepsilon_t \sim \mathcal{N}_n(\mathbf{0}, \Sigma_t), \\
h_{j,t} & = \mu_{h,j} + \phi_{h,j}(h_{j,t-1} - \mu_{h,j}) + \eta_t, \qquad \eta_t \sim \mathcal{N}(0,\sigma_{h,j}^2),
\end{align}
where $X_t = (\mathbf{1}_n, \, \mathbf{f}_t' \otimes I_k)$ is a $(n \times k_\beta)$ matrix with $k_\beta = n(k+1)$, $\bbeta = (\balpha_0', \, \operatorname{vec}(B)')'$ is an $k_\beta$-dimensional column vector, and $\Sigma_t = \operatorname{diag}(e^{h_{1,t}},\ldots,e^{h_{n,t}})$.
Moreover, we impose the stationarity constraint $\phi_{h,j} \in (-1,1)$ for each $j$.
An essential feature of this model is that it incorporates two types of risk exposure: exposure to the common factors, which can be considered the exogenous systematic risk, and exposure to the network linkages, which can be classified as endogenous systematic risk exposure.

In this specification, every network layer $W_i$ is associated with a weight parameter $\delta_i$ that controls the impact of the corresponding layer on the returns. Therefore, heterogeneous network reactions and multiple weighted layers can be considered simultaneously.
However, the multiplicative structure in the definition of $A_t$ complicates the separate identification of the layer-specific weights, $\bdelta = (\delta_1,\ldots,\delta_d)'$, and the country-specific network weights, $\brho$. Moreover, identifying the network-specific weights in $\bdelta$ becomes problematic as the correlation between the layers increases. As summarised in the following, we need to impose some restrictions on the parameter space to address this issue.

\begin{assumptions}
We make the following assumptions:
\begin{enumerate}[label=A\arabic*,font=\bfseries]
\item\label{hp:A1} $W_{i,t} \neq 0$ for each $j = 1,\ldots,d$ and $t=1,\ldots,T$.
\item\label{hp:A2} $W_{i,t} \neq W_{k,t}$ for each $i,k = 1,\ldots,d$, $i \neq k$, and $t=1,\ldots,T$.
\item\label{hp:A3} $\bdelta \in \Delta^{d-1}$, with $\Delta^{d-1}$ denoting the $(d-1)$-dimensional simplex. 
\item\label{hp:A4} $\operatorname{rank}(W_{k,t,\bullet}) > 0$ for each $k = 1,\ldots,n$ and $t = 1,\ldots,T$, where\\ $W_{k,t,\bullet} = \big[ W_{1,t}'|_{\bullet,k} \;\; W_{2,t}'|_{\bullet,k} \; \ldots  \; W_{d,t}'|_{\bullet,k} \big]$ and $W_{i,t}' |_{\bullet,k}$ denotes the $k$th row of $W_{i,t}$.
\end{enumerate}
\end{assumptions}

Assumption~\ref{hp:A1} ensures that at least one connection must exist within each network, eliminating the scenario of a network with no edges.
Assumption~\ref{hp:A2} states that two networks cannot be precisely identical; in that case, the networks cannot be identified correctly. If this occurs, one of the identical networks must be excluded from the analysis.
By requiring $\bdelta \in \Delta^{d-1}$, Assumption~\ref{hp:A3} helps preventing that both $R$ and $\bdelta$ lead to the same composite network.
Assumption~\ref{hp:A4} is made to ensure the identifiability of $\brho$ in the presence of multilayer networks. To better interpret this requirement, let us denote with $W_t^*$ the composite multilayer network, that is:
\begin{equation}
W_t^* = \sum_{i=1}^d \delta_i W_{i,t}.
\label{eq:weightslayer}
\end{equation}
Then, Assumption~\ref{hp:A4} imposes that every row of $W_t^*$ should contain at least one non-null element. Would this not hold, for instance, because the $k$th row of $W_{i,t}$ is always zero, then the associated $\rho_k$ would not be identifiable.

The composite matrix $W_t^*$ formed out of summing the separate layers is advantageous for interpretation and will match this condition when the individual networks are all row-normalised due to the abovementioned assumptions.
%
Row-normalising each $W_t$ in the standard way might diminish the effect of network density changes over time.
To further elaborate hereon, one could, for instance, take a situation into account where in a binary network, at time $t$ agent $i$ is connected solely to agent $j$ (with $w_{i,j}$ = 1 since a connection between these agents exists). Now, suppose that at time $t+1$, this connection remains perfectly unchanged, but country $i$ is not linked solely to agent $j$ anymore but also to multiple other assets. A general row-normalisation procedure would impose the entry $\frac{1}{m}$ for $w_{i,j}$ now, whereas $m$ is the total number of other agents that agent $i$ is linked to. This might cause improper analyses since a change on the entry $w_{i,j}$ is implied, but the relationships have not changed in any form at all.
To overcome this matter, a new approach called \textit{max-row normalisation} is defined as:
\begin{equation}
W_{i,j,t} = W_{i,j,t}^{U}\left(\max_{t} \sum_{i=1}^{N} W_{i,j,t}^{U}\right)^{-1},
\label{maxrow}
\end{equation}

With Assumptions \ref{hp:A1} to \ref{hp:A4}, some other restrictions are required for the identification and stability of the model. For instance, the non-singularity of $A_t$ is a necessary condition for the existence of a reduced form representation of the model in eq.~\eqref{eq:model_structural}.
For spatial autoregressive models that consider a single weight parameter, $\rho$, and one row-standardised static spatial matrix, $W$, Lemma 2 in \cite{sun1999posterior} provides a restricted region of support for the parameter $\rho$ to ensure (spatial) stationarity of the model. Specifically, they show that $\rho$ must lie in the interval $\lambda_{min}^{-1} < \rho < \lambda_{max}^{-1}$, where $\lambda_{min},\lambda_{max}$ denote the minimum and maximum eigenvalues of $W$.
%
Instead, in our more general time-varying multilayer network case ($d > 1$), no simple condition has been found to ensure that $A_t$ is invertible.
From now on, we apply row-normalisation to the matrices $W_{i,t}$, such that each row sums to 1 for each layer and time, and impose the restriction $\rho_j \in (-1,1)$ for each $j=1,\ldots,n$. This restriction corresponds to the effective region of support for the spatial weight parameter in the SAR model with a single-layer static network and single weight.
For completeness, we note that $\rho_j = 0$ for each $j$ makes it impossible to distinguish between alternative connectivity, meaning that the parameters $\bdelta$ are not identifiable. However, using an absolutely continuous prior and the information from the data rule out this extreme case in all our studies on synthetic and real data.

\begin{remark}
Recently, \cite{bonaccolto2019estimation} proposed an extension of the SAR model in eq.~\eqref{eq:model_SAR} by simultaneously including multiple layers and adopting a maximum likelihood approach to estimation. However, they focused on time-invariant spatial weight matrices, defined as pairwise Granger-causal networks obtained from a preliminary analysis of financial data.
In related work, \cite{debarsy2022bayesian} investigated the role of multiple weight matrices in a Bayesian SAR model and developed a simulation-based method to compute the marginal likelihood for model averaging purposes.
We differ from both approaches as we investigate a panel of countries with country-specific network exposure and time-varying multilayer networks. Moreover, we adopt a Bayesian approach to inference that allows for uncertainty quantification about the key parameters but relies on a different prior specification and computation approach than \cite{debarsy2022bayesian}.
\end{remark}

To get further insights into the implications of introducing time-varying spatial weight matrices, consider the reduced-form representation of the model in Eq.~\eqref{eq:model_structural}:
\begin{equation}
\label{eq:model_reduced}
\begin{split}
\by_t & = A_t^{-1} \balpha_0 + A_t^{-1} B \mathbf{f}_t + A_t^{-1} \bepsilon_t, \qquad \bepsilon_t \sim \mathcal{N}_n(\mathbf{0}, \Sigma_t) \\
 & = X_t \bbeta_t^* + \bepsilon_t^*, \qquad \bepsilon_t^* \sim \mathcal{N}_n(\mathbf{0}, A_t^{-1} \Sigma_t A_t^{-1}).
\end{split}
\end{equation}
Including time-varying weight matrices reduces form parameters that change over time, although the structural parameters are constant.
%
Moreover, under if Assumptions \ref{hp:A1} to \ref{hp:A4} hold, then $A_t^{-1}$ can be expanded as:
\begin{equation}
A_t^{-1} = (I_n - R W_t^*)^{-1} = I_n + \sum_{\ell=1}^\infty (R W_t^*)^\ell,
\label{eq:SAR_expansion}
\end{equation}
where each $\ell$th component of the sum represents the spatial effects of order $\ell$. Plugging the expansion of eq. \eqref{eq:SAR_expansion} into eq. \eqref{eq:model_reduced} yields:
\begin{equation}
\by_t = \sum_{\ell=1}^\infty (R W_t^*)^\ell X_t \bbeta + \sum_{\ell=1}^\infty (R W_t^*)^\ell \bepsilon_t + X_t \bbeta + \bepsilon_t.
\label{eq:model_reduced_SAR_expansion}
\end{equation}
The first two terms capture the indirect effects of the covariates and innovation stemming from the cross-sectional dependence encoded by the network structure in $W$. In contrast, the last two terms report the direct effects.
The expansion in eq.~\eqref{eq:model_reduced_SAR_expansion} shows that even though $B_{i,m} = 0$, the $m$th common factor can still affect the $i$th subject. For instance, if $B_{k,m} \neq 0$ and $W_{i,k} \neq 0$, meaning that subject $i$ is connected to subject $k$ and the latter is affected by the $m$th common factor, then subject $i$ will be indirectly affected by the same factor, through the connection encoded by the network. Conversely, by imposing $W=\mathbf{0}_n$ in eq.~\eqref{eq:model_SAR}, the $m$th common factor is irrelevant to the $i$th response whenever $B_{i,m} = 0$ in the resulting linear model.

\begin{remark}
A useful specification for forecasting purposes considers lags of the exogenous variables and networks. In this case, eq.~\eqref{eq:model_structural} would become:
\begin{align}
\label{eq:model_structural_dynamic}
\notag
A_{t-1} \by_t & = X_{t-1} \bbeta + \bepsilon_t, \qquad \bepsilon_t \sim \mathcal{N}_n(\mathbf{0}, \Sigma_t), \\
h_{j,t} & = \mu_{h,j} + \phi_{h,j}(h_{j,t-1} - \mu_{h,j}) + \eta_t, \qquad \eta_t \sim \mathcal{N}(0,\sigma_{h,j}^2).
\end{align}
\end{remark}

\subsection{Spillover effects}

To analyse the impact of shocks on the response variables, it is helpful to rewrite the model in its reduced form as
\begin{equation}
    \by_t = (I_n - R W_t^*)^{-1} (X_t \bbeta + \bepsilon_t),
\end{equation}
where the $n\times n$ matrix $S_t \coloneqq (I_n - R W_t^*)^{-1} = I_n + R W_t^* + (R W_t^*)^2 + \cdots$ is commonly referred to as the spatial multiplier matrix. It describes the transmission of demand shocks $\bepsilon_t$ (and thus of the elements in $X_t$) in the system of cross-response variable dependencies. The elements on the diagonal of the spatial multiplier matrix $S_t$, called direct effects in the spatial econometrics literature \citep[e.g., see][]{lesage2009introduction}, contain information about the impact of a shock $\epsilon_{i,t}$ on the variable $y_i$, namely the own effect.
The series expansion of the spatial multiplier matrix shows that these main diagonal elements already contain feedback effects due to cross-variable dependencies. The off-diagonal elements in $S_t$ comprise the indirect (or spillover) effects and capture the impact of a shock $\epsilon_{i,t}$ on other variables $y_j$, $\forall\ j\neq i$. 

Let us define two $n$-dimensional vectors of direct and indirect effects.
The $n$-dimensional vector $\bgamma_t$ collects the \textit{direct effects} at time $t$, meaning that the $i$th element $\gamma_{i,t}$ captures the direct effect of a shock to the $i$-th response variable.
Instead, the contribution of the \textit{indirect effects} (or spillovers), $\bzeta_t$, to the response variables is given by the off-diagonal elements of the spatial multiplier and measures the impact of shocks to a particular variable on each response resulting only from the interaction with other variables. In formulas, one has
\begin{align}
    \label{eq:direct_effects}
    \bgamma_t & = \biota'_n S_t^d, \\
    \label{eq:indirect_effects}
    \bzeta_t & = \biota_n' \big( S_t - S_t^d \big),
\end{align}
where $S_t^d$ is a diagonal matrix containing the diagonal elements of $S_t$.
Finally, the \textit{total effect} combines both direct and spillover effects and is given by 
\begin{equation}
\btau_t = \bgamma_t + \bzeta_t = \biota_n' S_t.
\label{eq:total_effects}
\end{equation}


\section{Bayesian Inference}  \label{sec:inference}

Let $\bsigma^2 = (\sigma_1^2,\ldots,\sigma_n^2)'$ and define with $\btheta = \{ \bbeta, \bsigma^2, \bdelta, \brho \}$ the collection of all parameters and $\by = \{ \by_1,\ldots,\by_T \}$ be the collection of all observations. To obtain the likelihood function, we define $\by_t^* = A_t \by_t$ in the structural form~\eqref{eq:model_structural} as in \cite{lesage2009introduction}, such that the inverse transformation has Jacobian $\abs{A_t}$, leading to:
\begin{equation}
\begin{split}
L(\by | \btheta) & = (2\pi)^{-\frac{nT}{2}} \abs{\Sigma}^{-\frac{T}{2}} \prod_{t=1}^T \abs{A_t} \exp\bigg\{\!\! -\frac{1}{2}(A_t \by_t - X_t\bbeta)' \Sigma^{-1} (A_t \by_t - X_t\bbeta) \bigg\}.
\end{split}
\label{eq:likelihood}
\end{equation}

\subsection{Prior specification}
We assume a Gaussian prior distribution for the coefficient vector $\bbeta$, which encodes the exposure to the common factors. However, in high-dimensional settings, this choice can be easily generalised to global-local shrinkage priors \citep[e.g., see][]{carvalho2010horseshoe,bhattacharya2015dirichlet} for sparse estimation of the coefficients, without substantial change in the computational cost.
Concerning the variance, we assume an improper prior on the country-specific log-variance, which results in a flat prior on $\sigma_j^2$ \citep[see also][]{dittrich2017bayesian}, $p(\sigma_j^2) \propto \sigma_j^{-2}$. For adequately large samples and given the functional form of the likelihood, the full conditional posterior is proper, and this prior is not influential on posterior estimates.

Following the discussion in Section~\ref{sec:model}, the layer-specific weight parameters $\bdelta$ are constrained to lie in the $(d-1)$-dimensional simplex. Therefore, we assume a Dirichlet prior distribution to impose such restrictions.
Finally, as each country-specific network weight $\rho_j$ on the diagonal of the matrix $R$ is restricted to the interval $(-1,1)$, we enforce such constraint by assuming a Beta prior on $\tilde{\rho}_j = (\rho_j+1)/2$ \citep[see also][]{kastner2014ancillarity}.
See \cite{lesage2007bayesian} for a discussion on alternative prior distributions for $\rho_j$.

The set of all prior distributions for the parameters of the model is:
\begin{equation}
\begin{aligned}
\bbeta & \sim \mathcal{N}_{k_\beta}(\bbeta | \underline{\bmu}_\beta, \underline{\Sigma}_\beta),
 & \sigma_j^2 & \propto 1/\sigma_j^2 \\ 
\bdelta & \sim \mathcal{D}ir(\bdelta | \underline{\bc}), 
 & \qquad \frac{\rho_j+1}{2} & \sim \mathcal{B}e\Big( \frac{\rho_j+1}{2} \Big| \underline{a}_\rho, \underline{b}_\rho \Big).
\end{aligned}
\label{eq:priors}
\end{equation}

\subsection{Posterior Sampling}

Let $\brho_{-j}$ and $\bsigma_{-j}^2$ denote all the elements of $\brho$ and $\bsigma^2$, respectively, except the $j$th.
To draw samples from the posterior distributions, we design an MCMC algorithm that is structured as follows:
\begin{enumerate}[label=\arabic*)]
\item sample $\bbeta | \by, \bsigma^2, \brho, \bdelta$ from a  Gaussian distribution;
\item sample $\sigma_j^2 | \by, \bsigma_{-j}^2, \bbeta, \brho, \bdelta$ from an inverse gamma distribution;
\item sample $\bdelta | \by, \bbeta, \bsigma^2, \brho$ from $P(\bdelta | \by, \bbeta, \bsigma^2, \brho)$ using an independent Metropolis-Hastings (iMH) algorithm;
\item sample $\rho_j | \by, \brho_{-j}, \bbeta, \bsigma^2, \bdelta$ from $P(\rho_j | \by, \brho_{-j}, \bbeta, \sigma_j^2, \bdelta)$ using a slice sampler algorithm \citep{neal2003slice}.
\end{enumerate}
We refer the reader to the Supplement for further details on the derivations of the posterior distributions.
The conditional distributions for $\bbeta$ and $\sigma_j^2$ are standard, and fast algorithms to sample from them are used. In high-dimensional set-ups with numerous covariates, the introduction of global-local shrinkage priors for $\bbeta$ based on scale mixtures of Gaussian distributions \citep[e.g., see][]{polson2010shrink,bhadra2016default} does not affect significantly the computational cost, as only draws from standard distributions are required.

The main computational challenge concerns sampling the layer-specific weight parameters, $\bdelta$, and the network weight parameter, $\rho_j$ for any $j=1,\ldots,n$, since there is no distribution conjugate with the likelihood in eq.~\eqref{eq:likelihood}. Besides, the constraints discussed in Section~\ref{sec:model} give an additional layer of complexity.
Adopting a full Bayesian approach to inference, we introduce the restrictions on the parameters $\bdelta$ and $\brho$ through appropriate choice of the prior distributions in eq.~\eqref{eq:priors}, which lead to non-standard posterior full conditional distributions.

Therefore, we individually sample each $\rho_j$ from its full conditional distribution using a slice sampler algorithm \citep{neal2003slice}.
This approach allows us to sample from an arbitrary univariate distribution (known up to a proportionality constant) via introducing an auxiliary (slice) variable. Specifically, denoting by $f$ the full conditional distribution, of $\rho_j$ and $\rho_j^{(m)}$ the value of $\rho_j$ at the $m$th iteration, we obtain a draw of $\rho_j^{(m+1)}$ as
\begin{equation}
\begin{split}
u_j^{(m)} | \rho_j^{(m)} & \sim \mathcal{U}\big( 0, f(\rho_j^{(m)}) \big), \\
\rho_j^{(m+1)} | u_j^{(m)} & \sim \mathcal{U}\big( \mathbb{A}^{(m+1)} \big), \qquad \mathbb{A}^{(m+1)} = \big\{ x : f(x) \geq u_j^{(m)} \big\}.
\end{split}
\end{equation}
By relying on sampling from univariate full conditional distributions, this approach does not suffer from the tuning issue and potential low acceptance rate of Metropolis-Hastings techniques more traditionally used.
Also, any draw from this distribution belongs to the interval $(-1,1)$ by the (transformed) beta prior in eq.~\eqref{eq:priors} that incorporates this restriction.

Concerning the network-specific weights given by the vector $\bdelta$, we make sure that the constraint in \ref{hp:A3} is satisfied by assuming a Dirichlet prior in eq.~\eqref{eq:priors}, then jointly updating the entire vector at each MCMC iteration. Specifically, we design an independent Metropolis-Hastings (iMH) algorithm with a Dirichlet proposal distribution and use a preliminary short run of the MCMC algorithm to tune the hyperparameter.
The proposed method works well in both simulation studies and the real-data application. As the number of network layers $d$ increases, the acceptance rate of the iMH algorithm tends to decay, suggesting the use of entry-wise updates \citep[see][for a possible approach]{debarsy2022bayesian}.

\section{Application}  \label{sec:application}
We consider the Group of Seven (G7) economies, comprising Canada (CA), France (FR), Germany (DE), Italy (IT), Japan (JP), the United Kingdom (UK), and the United States (US).
The G7 is a coalition of developed global economies collaborating to devise unified strategies for tackling significant worldwide issues. 
These encompass domains like international trade, security concerns, and climate change. 
The G7 group encompassed 30.39\% of the world GDP and, as of January 2023, it collectively accounted for approximately 77\% of the total global equity market value, indicating their significant presence in the world's largest stock markets.\footnote{Data have been retrieved at the International Monetary Fund \url{https://www.imf.org/external/datamapper/profile/MAE} and from stock market data.}
The relationships within the G7 can exhibit both positive and negative dynamics through various channels, such as trade agreements, geopolitical events, and economic policies. 
During tranquil periods, often indicative of economic cooperation or shared interests, positive relationships can boost investor confidence and stock market indices. 
Conversely, trade disputes that may result in trade barriers, political tensions, or economic crises represent hostile relations that can influence market volatility and lead to a depreciation in stock market performance.
Therefore, incorporating the dynamics of these relationships as a multilayer network for positive and negative relationships might provide a valuable source of information for explaining the performance of the respective financial markets.
This article presents two applications involving G7 stock market returns and volatilities over the sample period under consideration.

\subsection{Data description}
Our analysis examines the connection between the G7 countries, which constitute the multilayer network, along with business and consumer indicators as control variables, and their impact on the stock market indices. The timeframe considered for this analysis extends from January 1998 to December 2022, with data sampled monthly.

\paragraph{Country relationship data}
Relationship data among G7 countries is obtained from the Integrated Crisis Early Warning System (ICEWS) database.\footnote{The ICEWS database is accessible for download at \url{https://dataverse.harvard.edu/dataset.xhtml?persistentId=doi:10.7910/DVN/28075}. It is worth noting that ICEWS was discontinued in April 2023 and has been succeeded by the POLECAT database.}
The data encompass encoded interactions among socio-political entities involving cooperative or adversarial actions spanning individuals, groups, sectors, and countries.
ICEWS involves articles and various news from news outlets and media platforms where the initial six sentences are processed into structured information in the form of ``events''.
Each event encompasses details about the originating and receiving actors and the category identified by the CAMEO event taxonomy developed by \cite{gerner2002conflict}, containing approximately 312 distinct categorical event types.
Examples of positive events include ``Cooperate Economically'', Grant Diplomatic Recognition'', ``Ease Economic Sanctions, Boycott, Embargo'', and ``Provide Military Protection or Peacekeeping''. 
Conversely, adverse events encompass ``Halt negotiations'' ``Reduce or Break Diplomatic Relations'' ``Impose Embargo, Boycott, or Sanctions'', and ``Expel or Deport Individuals''. 
The ICEWS database provides information about the source and target countries and specifies each event's main actors or groups.
Each event is assigned an ``intensity'' value within the interval $[-10,10]$, signifying the event's impact on the relationship between the two actors.
For instance, events characterised by highly negative intensities, like ``Use conventional military force'' (intensity $-10$), indicate a strong weakening of the relationship. In contrast, events with positive intensities, such as ``Retreat or surrender militarily'' (intensity $10$), suggest a potential convergence of interests between the two agents. 

Table~\ref{tab:event_ex} provides an illustrative example of a positive and negative event from the ICEWS database. 
For instance, on 4 January 1998, an event with CAMEO code 111, indicating the signing of a formal agreement, occurred with an intensity of 8. The source country was Japan, and the target was the United States. 
On the following day, a different event took place with a CAMEO code of 172, signifying the imposition of administrative sanctions. This event had an intensity of -5 and involved France as the source country and the United Kingdom as the target country. 
We employ the positive and negative domains of event intensity to delineate cooperative and conflictual relationships, forming the basis for constructing the two distinct layers within the network. 
This approach provides a nuanced understanding of international interactions, allowing us to capture both collaborative efforts and contentious encounters. 
Alternatively, one might consider using the net intensity value to encapsulate an event's overall impact. 
However, by preserving the positive and negative components separately, we can account for the inherent asymmetry in the effects exerted by these distinct dimensions on the stock markets. 
In our view, this approach enhances the granularity of our network model and enables a more comprehensive evaluation of cross-country dynamics.
In our analysis, we exclude domestic events\footnote{We define domestic events as those where the source and target country are the same.} since our focus is on modelling cross-country relationships, where the source and target countries are distinct. The total number of observed events during the period is 824,703, which reduces to 348,334 when excluding domestic occurrences.

\begin{table}[t!h]
\centering
\resizebox{\textwidth}{!}{
\begin{tabular}{c c c c c c}
\hline
Date & CAMEO & Event & Intensity & Source Country & Target Country  \\
\hline
04-01-1998 & 111 & Sign formal agreement & 8 & Japan & United States \\
05-01-1998 & 172 & Impose administrative sanctions & -5 & France & United Kingdom
\end{tabular}
}
\caption{An illustrative example of a cooperative and conflictual event from the ICEWS database.}
\label{tab:event_ex}
\end{table}

The network consists of two time-varying distinct layers where the entries, denoted as $w_{i,j,t}^{(l)}$, represent the cumulative intensities of all events originating from G7 country $i$ and directed towards G7 country $j$ at time $t$ in layer $l \in \{\textrm{cooperative},\textrm{conflictual}\}$. 
The directed nature of the network and the exclusion of self-loops imply that the adjacency matrix is asymmetric with zero diagonal elements.

\begin{figure}[t!h]
\centering
\captionsetup{width=0.95\linewidth}
\begin{tabular}{ccc}
    & cooperative layer & conflictual layer\\
    \begin{rotate}{90} \hspace*{70pt} November 1998 \end{rotate} \hspace*{-10pt} &
    \includegraphics[width=202pt,height=185pt]{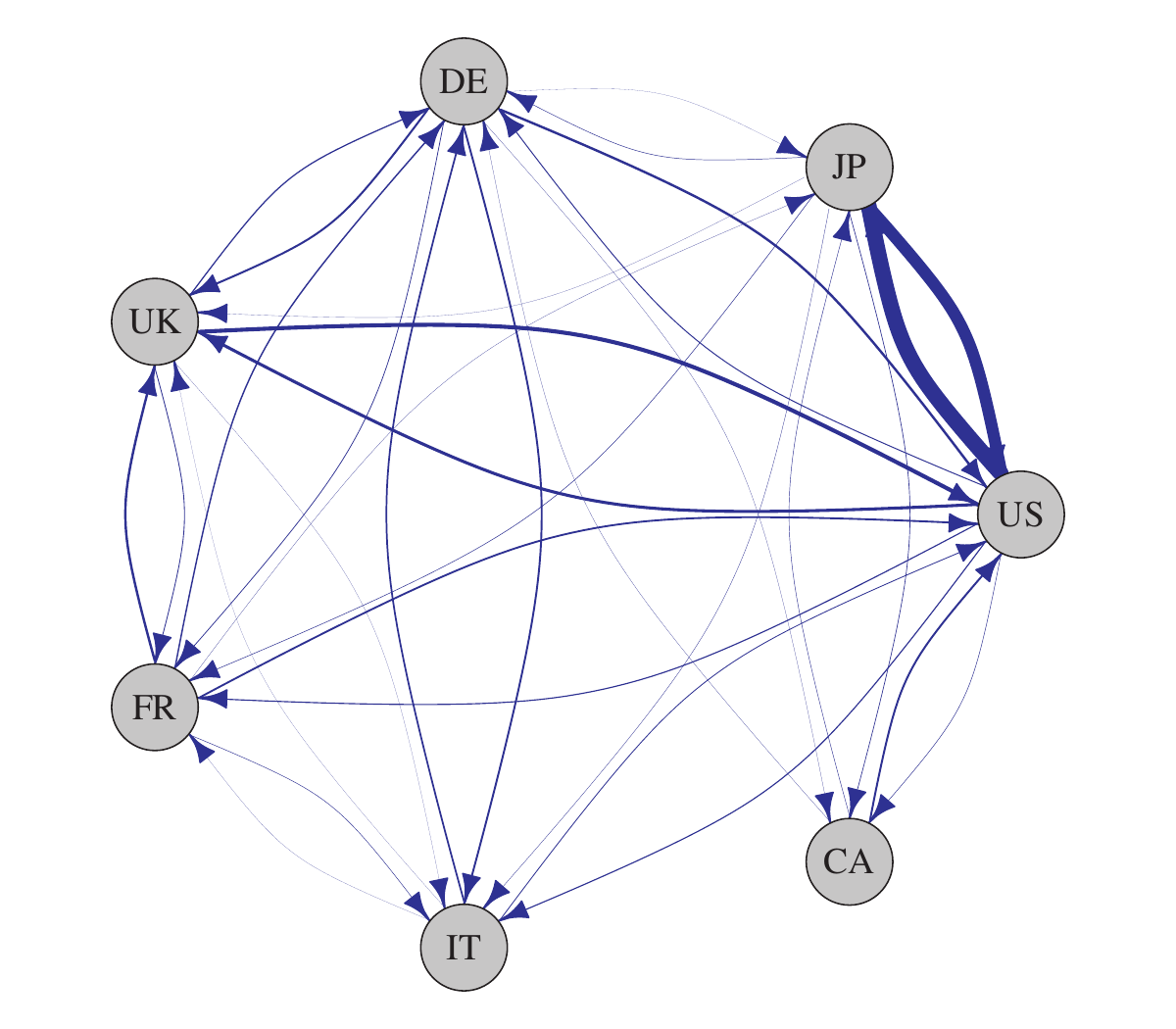} &
    \includegraphics[width=202pt,height=185pt]{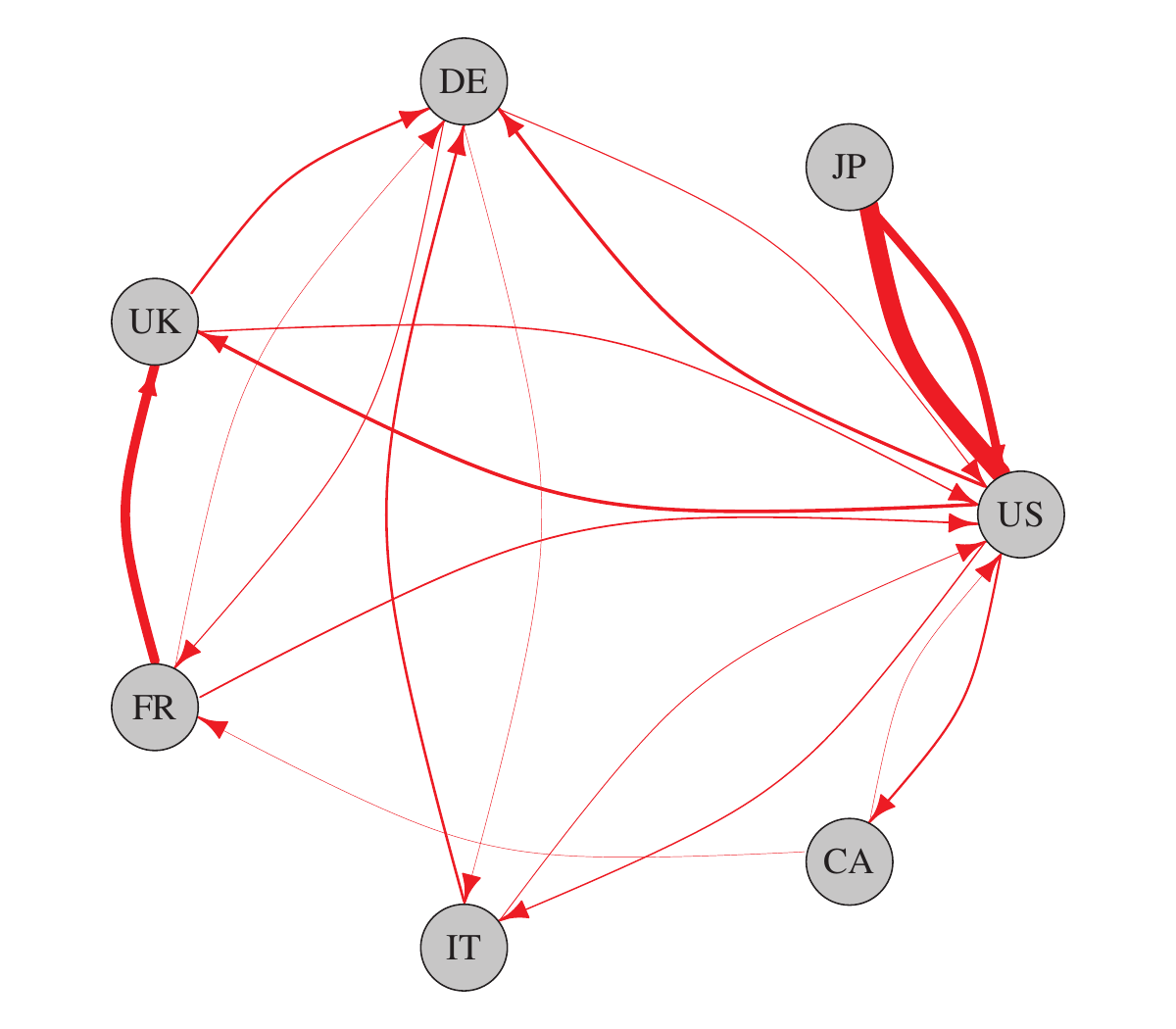} \\
    \begin{rotate}{90} \hspace*{70pt} March 2003 \end{rotate} \hspace*{-10pt} &
    \includegraphics[width=202pt,height=185pt]{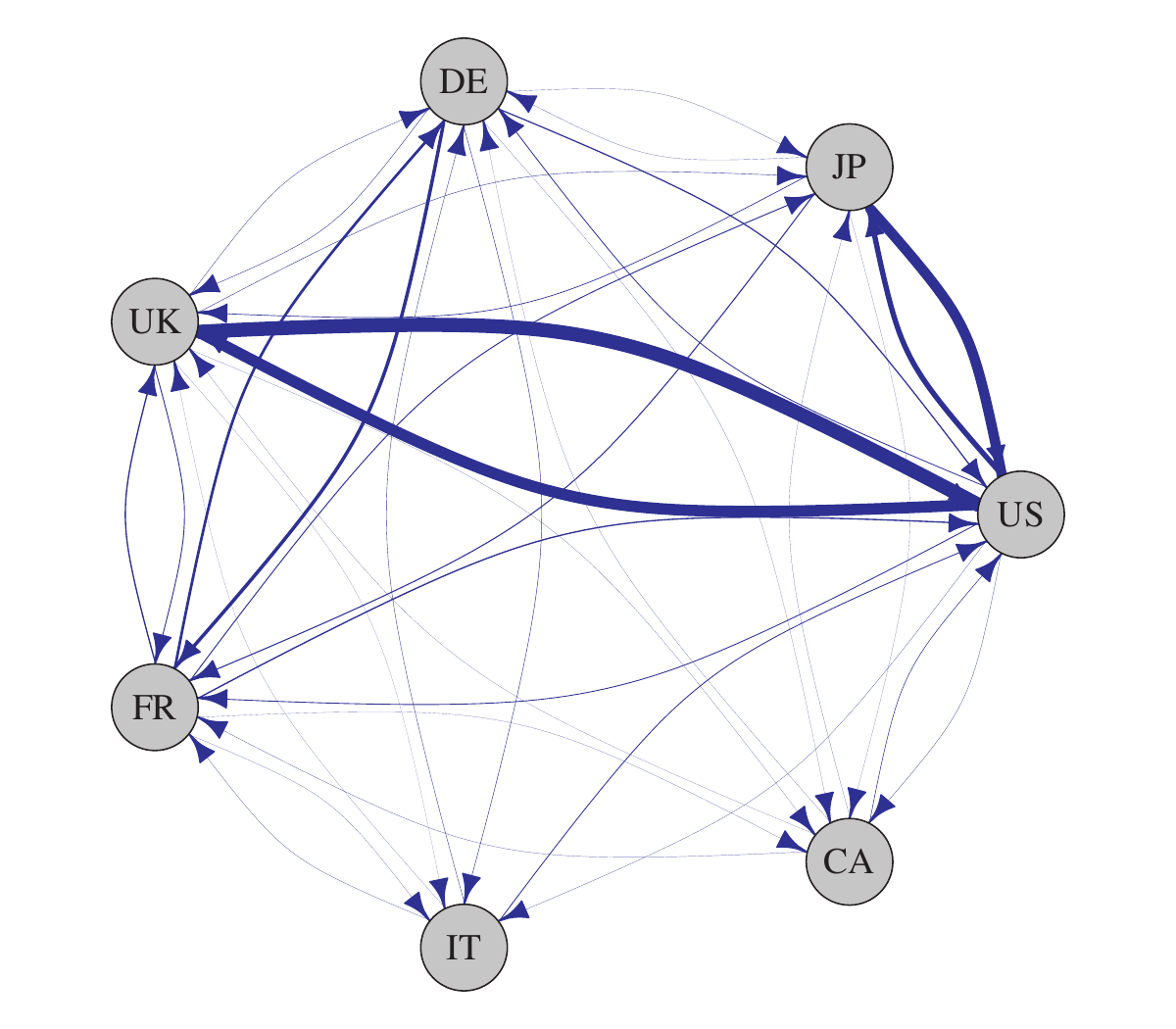} &
    \includegraphics[width=202pt,height=185pt]{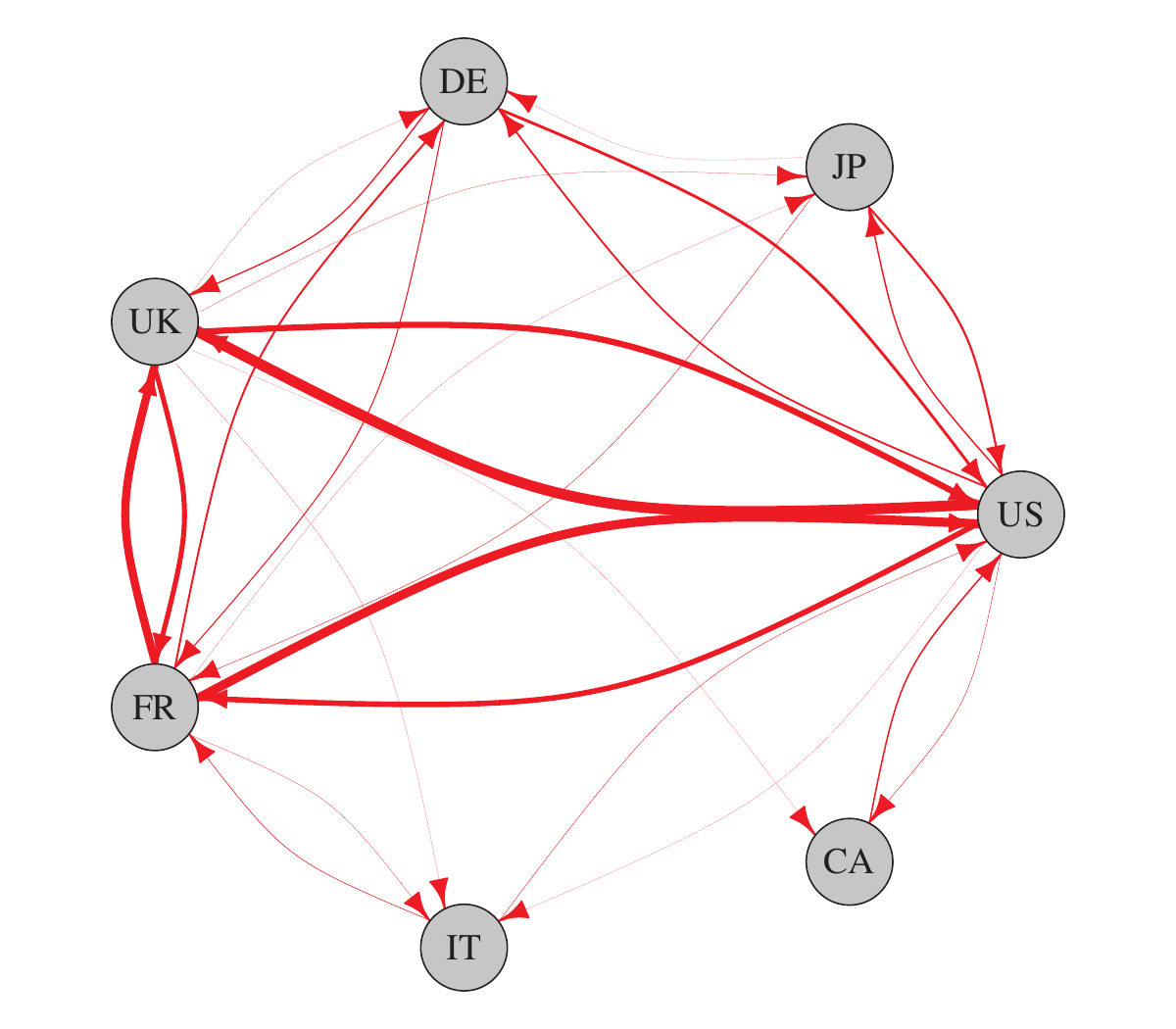} \\
    \begin{rotate}{90} \hspace*{70pt} May 2018 \end{rotate} \hspace*{-10pt} &
    \includegraphics[width=202pt,height=185pt]{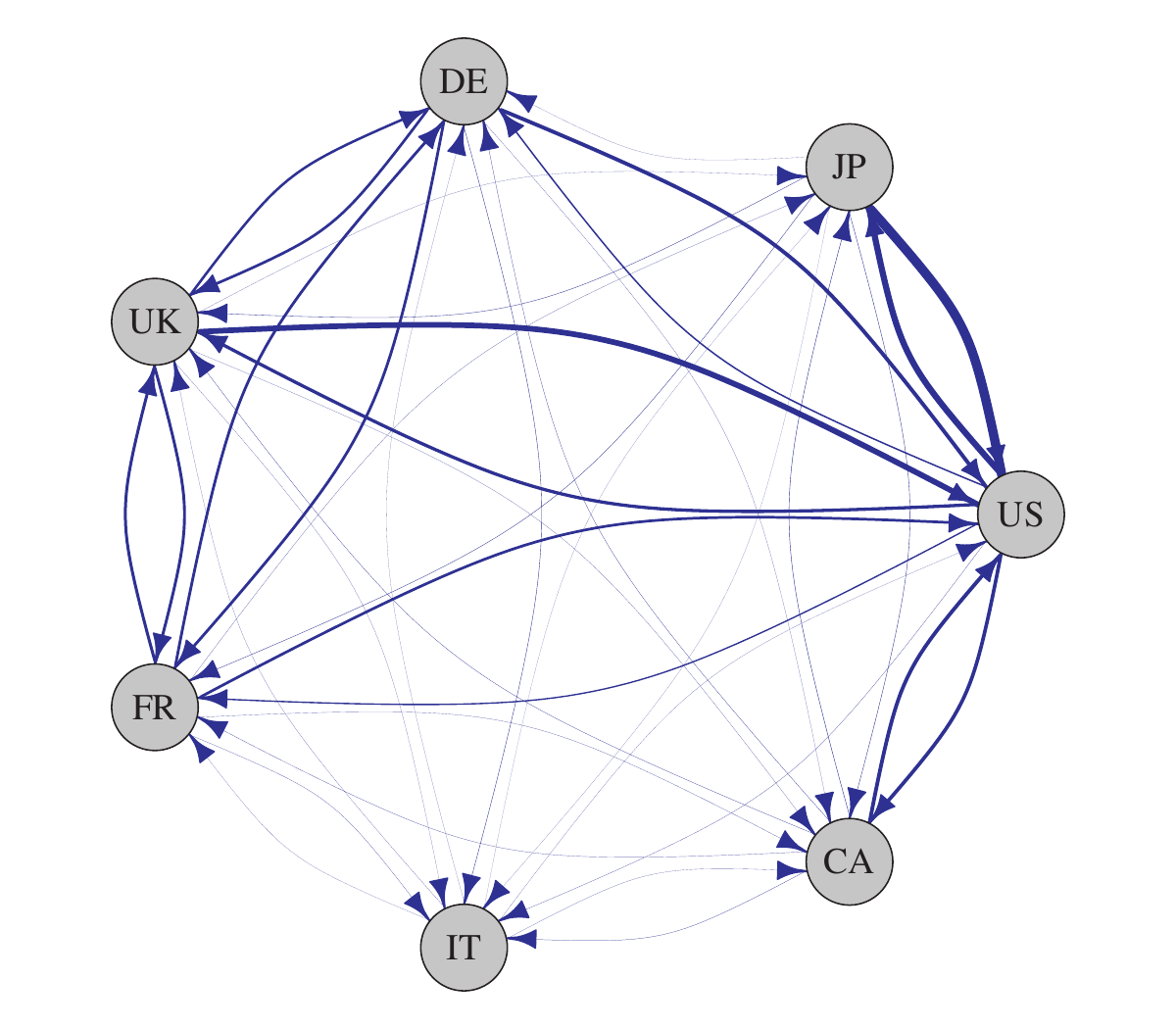} &
    \includegraphics[width=202pt,height=185pt]{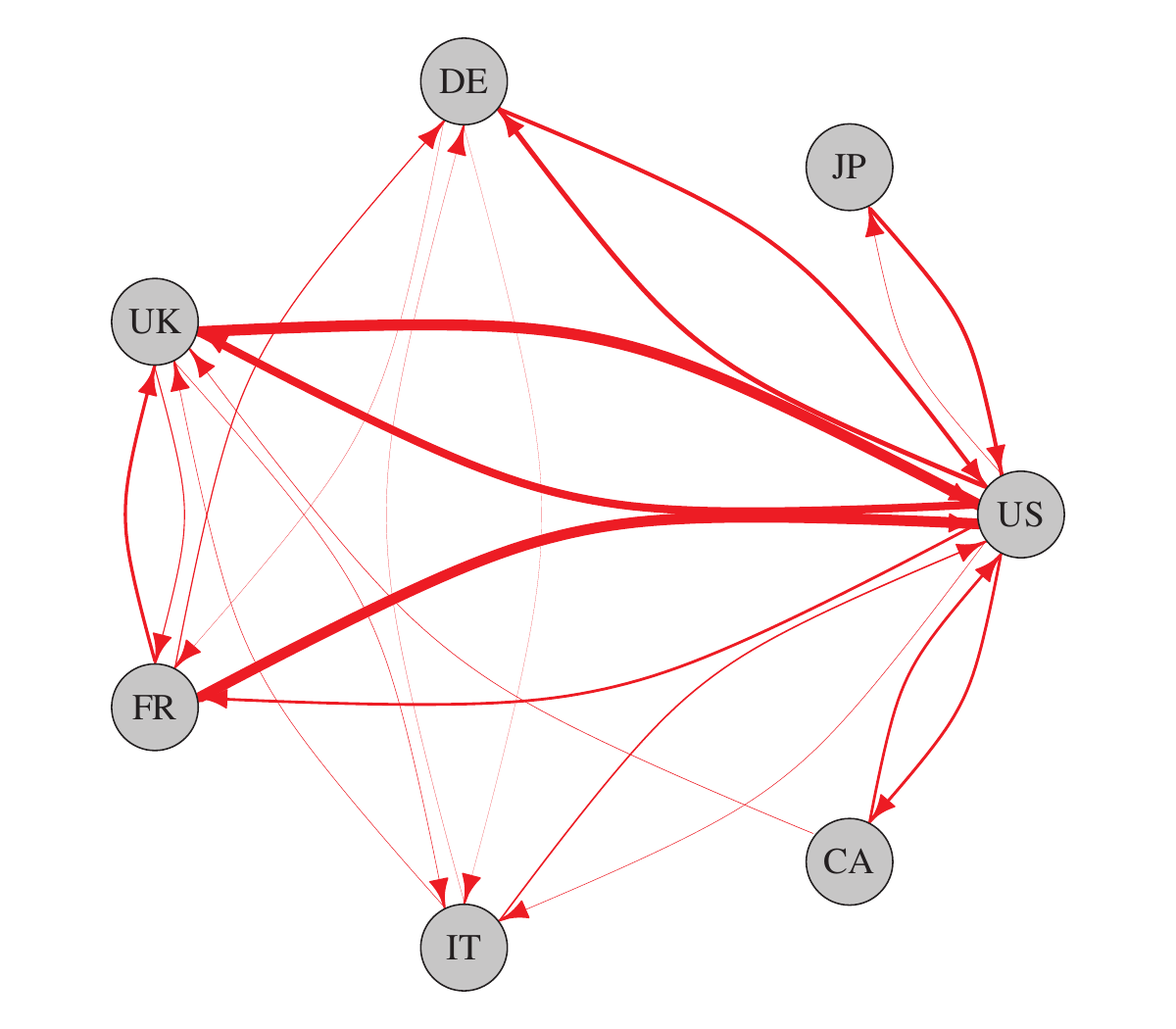} \\    
\end{tabular}
\caption{Network layers for cooperative interactions (blue edges, left column) and conflictual interactions (red edges, right column) on three specific dates: i) Japan-U.S. Summit in November 1998, ii) Invasion of Iraq in March 2003, and iii) Trade barriers imposed by the United States in May 2018 on Canada, the European Union, Mexico, and South Korea. Nodes are labelled with the country codes.}
\label{fig:net}
\end{figure}


As an illustrative example, Figure~\ref{fig:net} plots the network layers, distinguishing cooperative interactions in blue and conflictual interactions in red for three dates: the Japan-US Summit in November 1998, the Invasion of Iraq in March 2003, and the imposition of trade barriers by the United States in May 2018. 
These snapshots showcase different cooperative and conflict events.
On the first date (top panel), a symmetrical pattern emerges in the cooperative interactions (left column) between the United States and Japan. 
This resulted from the Japan-US Summit in November 1998, which involved substantial political, security, economic, and international discussions.\footnote{Detailed information can be found at \url{https://www.mofa.go.jp/region/n-america/us/visit98/summit.html}.}
These connections appear thicker compared to others. However, in the conflictual layer (right column), there is a connection from Japan to the US, which could indicate some disagreement within Japan or other unrelated events.
The second date (mid-panel) underscores the division prompted by the 2003 invasion of Iraq. The United States and the United Kingdom played pivotal roles in the intervention coalition, while France and Germany opposed it. A cooperative relationship existed between the US and the United Kingdom (left column), alongside conflictual episodes involving the United States, France, and the UK.
The last date (bottom panel) centres around the trade barriers imposed by the United States in May 2018, which led to conflicting relationships between the US, France, and the UK.

\begin{figure}[h!]
\centering
\small
\begin{tabular}{c c}
cooperative layer & conflictual layer \\
    \includegraphics[scale=0.20]{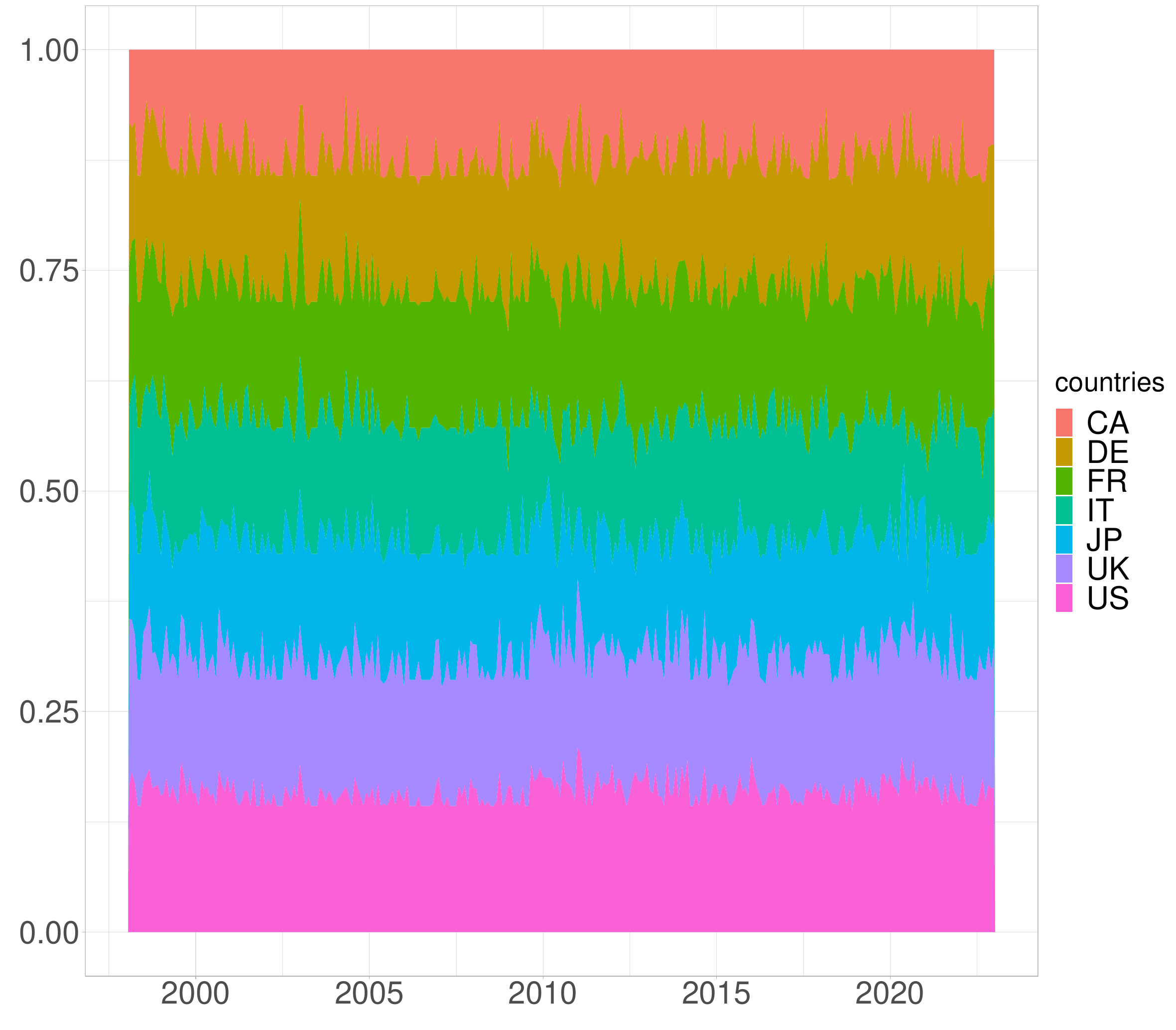} &
    \includegraphics[scale=0.20]{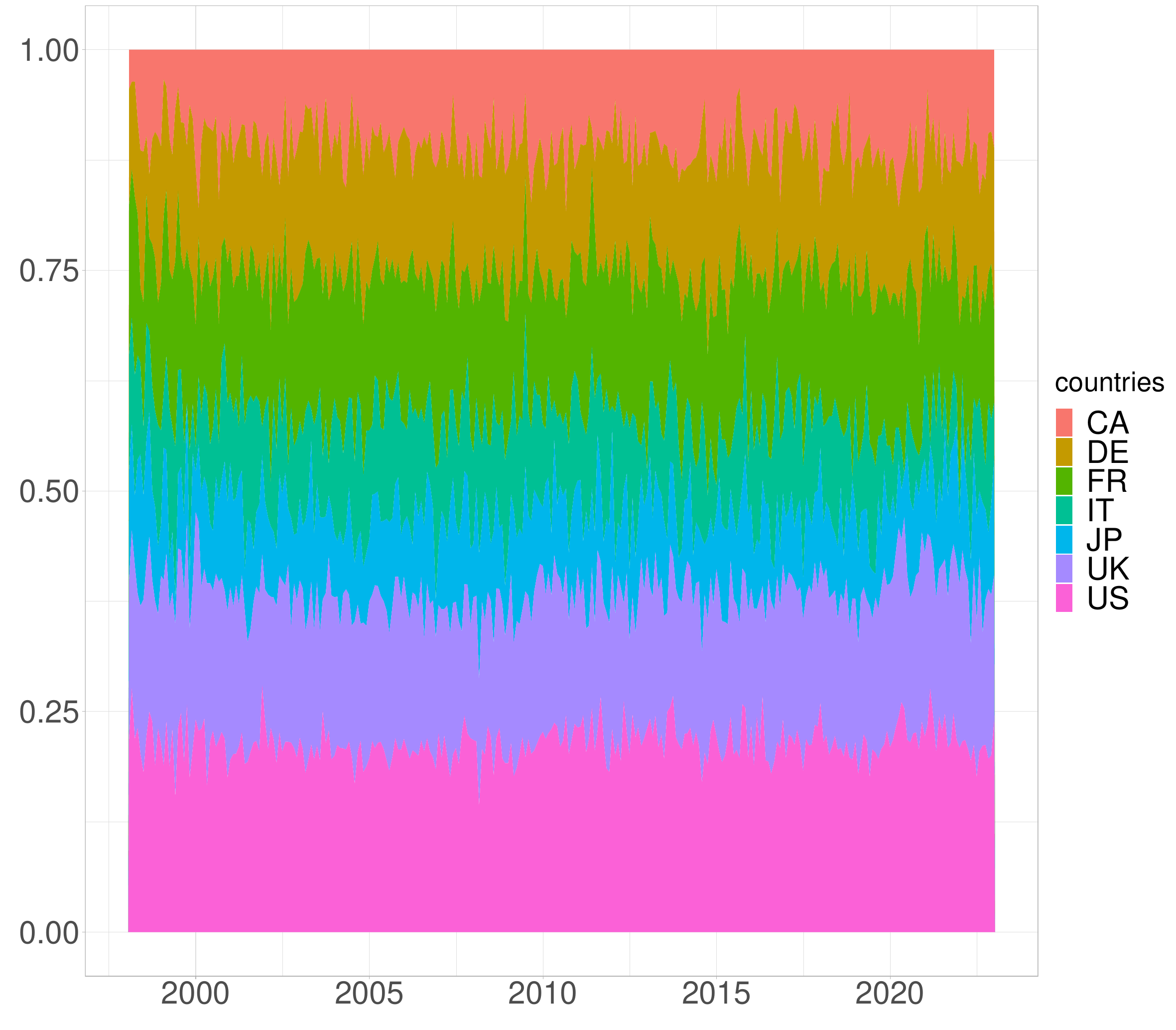}
\end{tabular}
\caption{Eigenvector centrality as a percentage of the total for G7 countries in the cooperative (left) and conflictual (right) layers over the entire period: Canada (CA, orange), France (FR, green), Germany (DE, gold), Italy (IT, light green), Japan (JP, blue), the United Kingdom (UK, purple), and the United States (US, magenta).}
\label{fig:eig}
\end{figure}

Figure~\ref{fig:eig} provides eigenvector centrality values for the G7 group in each network layer, measuring each country's relative influence and centrality over time.
Regarding both layers, the United States (magenta) consistently emerges as a pivotal player, especially in the conflictual one, exhibiting the highest centrality values throughout the observed period. 
Japan (blue) and Italy (light green) present a higher influence in the cooperative layer, while Germany (gold) and the United Kingdom (purple) demonstrate substantial centrality in both layers.
France (green), while influential, holds a position slightly subordinate to the nations mentioned above.
Notably, Canada (orange) displays the lowest centrality values, implying a relatively peripheral role in the examined cooperative relationships.

\FloatBarrier

\paragraph{Control variables}
As additional variables that could potentially influence stock market returns, we consider the OECD Business Confidence Index (BCI) and the OECD Consumer Confidence Index (CCI) for the G7 area.
The BCI offers insights into future trends by drawing from surveys on production, orders, and inventories of finished goods within the industrial sector.
In the second application, focusing on volatilities, we include the one-month lagged monthly realised volatility of the MSCI World Index as a control variable to account for volatility persistence in the financial markets.

\paragraph{Stock indices}
We collected price data for stock indices from Bloomberg for the following G7 countries: Canada (S\&P/TSX), France (CAC 40), Germany (DAX 30), Italy (FTSE MIB), Japan (Nikkei 225), the United Kingdom (FTSE 100), and the United States (S\&P 500). 
For the first application, we consider the monthly logarithmic returns, while for the second, we consider the realised volatility as follows:
\begin{equation}
    \mathrm{RV}_{i,t} = \sum_{l=1}^M r_{l,t}^2,
\end{equation}
where $i =$\{CA, FR, DE, IT, JP, UK, US\}, $t$ denotes the monthly time index, and $M$ represents the total number of market days in a given month $t$.
Additional information for the control variables and stock indices are reported in the Supplement.

\FloatBarrier
\subsection{Results}
This section presents the findings related to stock market returns and volatilities based on the model specification in eq.~\eqref{eq:model_reduced_SAR_expansion}.
%
Regarding the case of returns, we focus on the control variables in Table~\ref{tab:beta_static_control}, which presents the posterior mean along with the corresponding $95\%$ credible intervals for the coefficients $\bbeta$ associated with changes in the Business Confidence Index ($\Delta BCI$) and the Consumer Confidence Index ($\Delta CCI$) on the stock market returns of the G7 countries.
The statistically significant coefficients (i.e., zero outside the 95\% 
It is noteworthy that when estimating eq.~\eqref{eq:model_reduced_SAR_expansion} without considering the network layers, all controls show significance except for $\Delta BCI$ in the case of Italy (results have been included in the Supplement).
This highlights a spatial effect due to the collaborative and conflictual interactions among the G7 group.

\begin{table}[h!t]
\centering
\footnotesize
\begin{tabular}{cc c c c c c c c}
\toprule
 & \multirow{2}{*}{\textit{const}} & \multicolumn{7}{c}{$\Delta BCI$} \\
 & & US & JP & DE & UK & FR & IT & CA \\
\midrule
$\hat\beta$ & 0.0007 & \textbf{0.0442} & \textbf{0.0298} & -0.0098 & 0.0002 & -0.0014 & -0.0059 & \textbf{0.0251} \\ 
CI  2.5\% & -0.0003 & 0.0202 & 0.0021 & -0.0252 & -0.0136 & -0.0137 & -0.0249 & 0.0093 \\ 
CI 97.5\% & 0.0018 & 0.0699 & 0.0568 & 0.0057 & 0.0144 & 0.0120 & 0.0124 & 0.0418 \\
\midrule
 & & \multicolumn{7}{c}{$\Delta CCI$} \\
 & & US & JP & DE & UK & FR & IT & CA \\
\midrule
$\hat\beta$ &  & \textbf{0.0427} & -0.0008 & \textbf{0.0244} & 0.0014 & \textbf{0.0147} & \textbf{0.0317} & -0.0068 \\ 
CI  2.5\% &  & 0.0177 & -0.0289 & 0.0080 & -0.0142 & 0.0014 & 0.0148 & -0.0235 \\ 
CI 97.5\% &  & 0.0712 & 0.0268 & 0.0403 & 0.0161 & 0.0287 & 0.0497 & 0.0104 \\ 
\bottomrule
\end{tabular}
\caption{Posterior mean $\hat\beta$ and 95\% credible interval of the coefficients $\bbeta$ for G7 returns. Statistically significant coefficients (those with a 95\% credible interval excluding zero) are in bold.}
\label{tab:beta_static_control}
\end{table}

\begin{figure}[h!]
\centering
\includegraphics[scale=0.35]{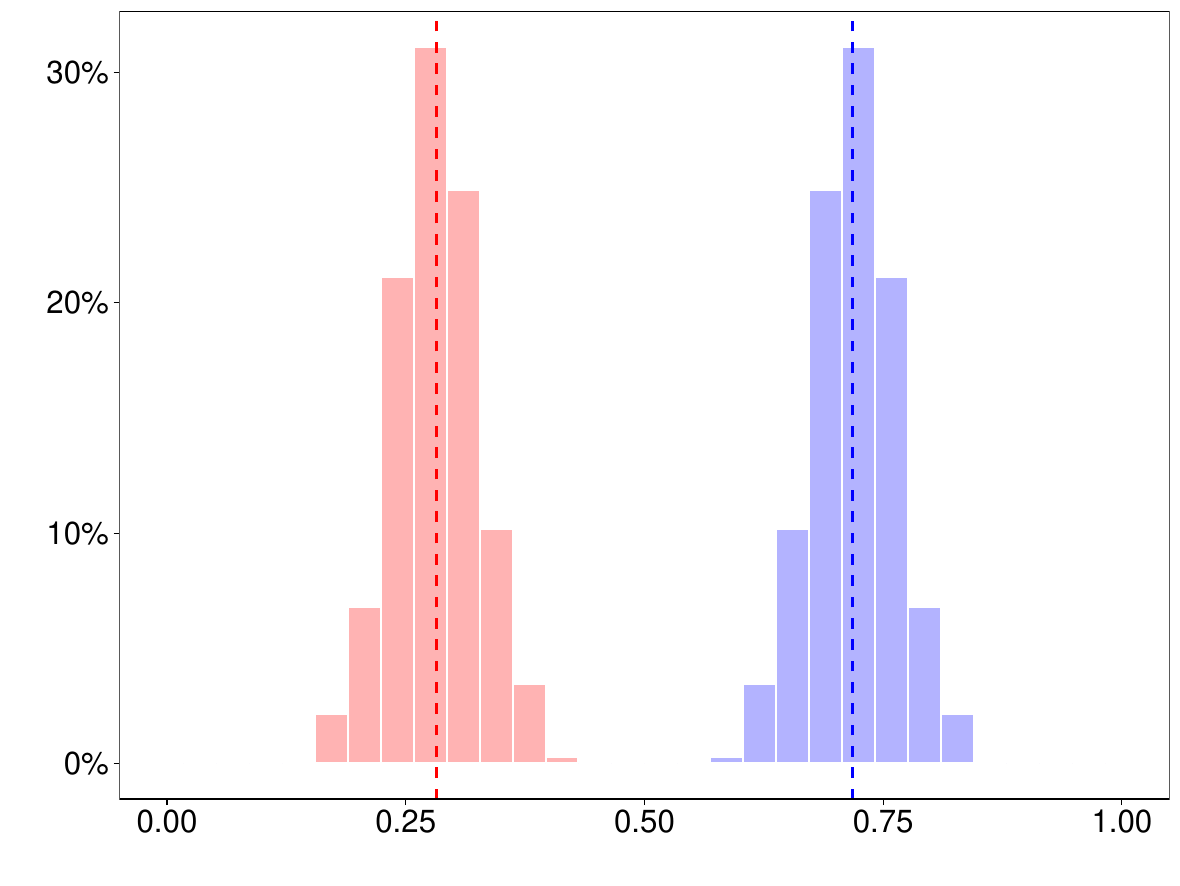}
\caption{Posterior distribution of $\bdelta$, with $\delta_1$ (weight of conflictual relationships layer) in red and $\delta_2$ (weight of cooperative relationships layer) in blue for G7 returns. The posterior mean is reported as a dashed line.
}
\label{fig:insample_delta}
\end{figure}

Next, we shift our attention to the layer-specific weights, as defined in eq.~\eqref{eq:At_def} and eq.~\eqref{eq:weightslayer}. 
This allows us to gauge the relative importance of each layer in influencing the network dynamics.
Figure~\ref{fig:insample_delta} plots the posterior distribution of $\delta_1$ (corresponding to the conflictual layer) and $\delta_2$ (representing the cooperative layer). Notably, the two distributions exhibit no overlap, thus providing evidence that these weights are distinct.
In particular, their respective contributions to the model differ noticeably, with the posterior mean of the cooperative layer ($\hat\delta_2 = 0.718$) surpassing that of the conflictual layer ($\hat\delta_1 = 0.282$).
This discrepancy in weights suggests a more substantial influence of cooperative interactions than conflictual ones on the stock market returns, emphasising the collaborative nature of the relationships within the G7 group.

\begin{figure}[h!]
\centering
\footnotesize
\begin{tabular}{c c c c}
US & JP & DE & UK \\
\includegraphics[scale=0.25]{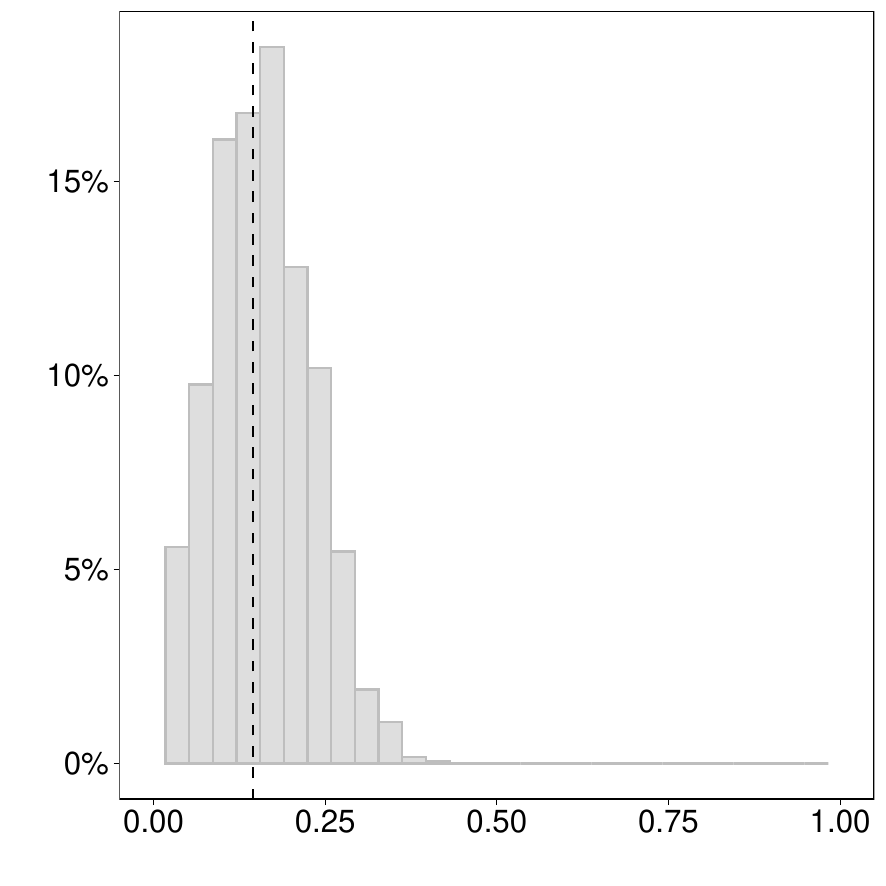} &
\includegraphics[scale=0.25]{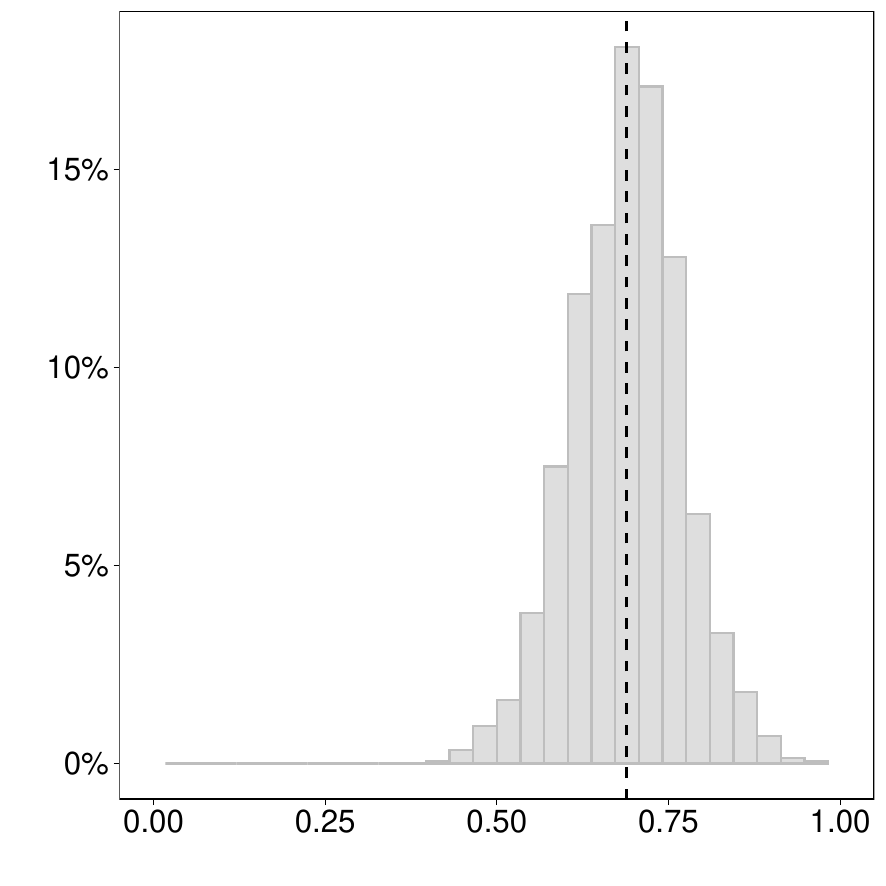} &
\includegraphics[scale=0.25]{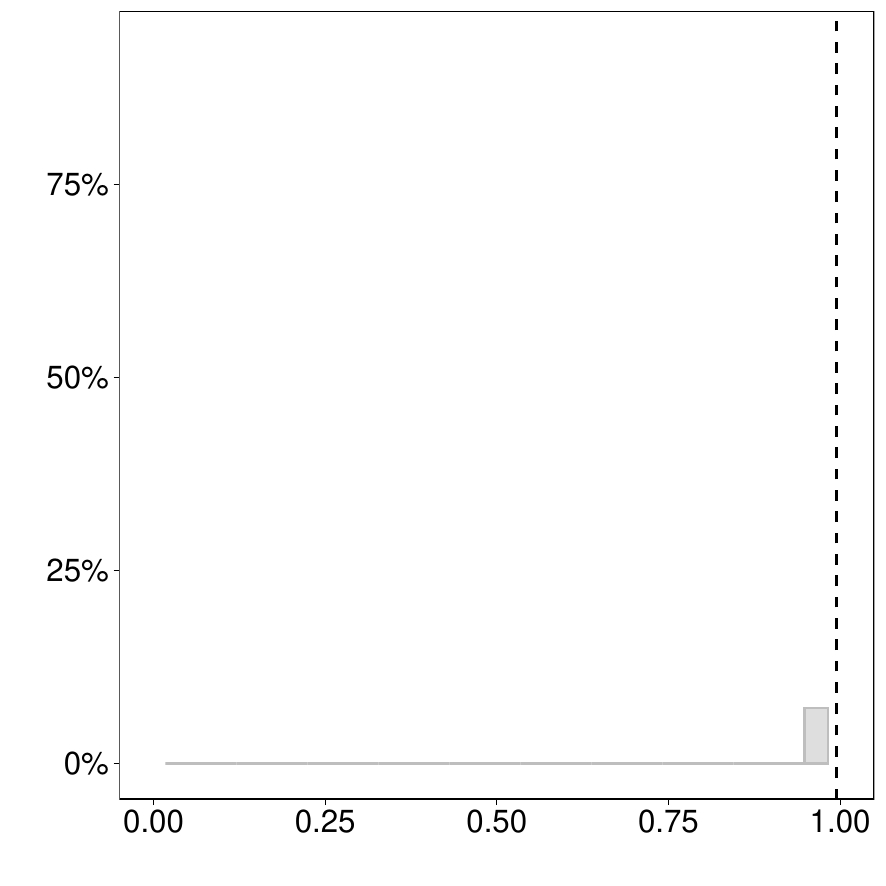} &
\includegraphics[scale=0.25]{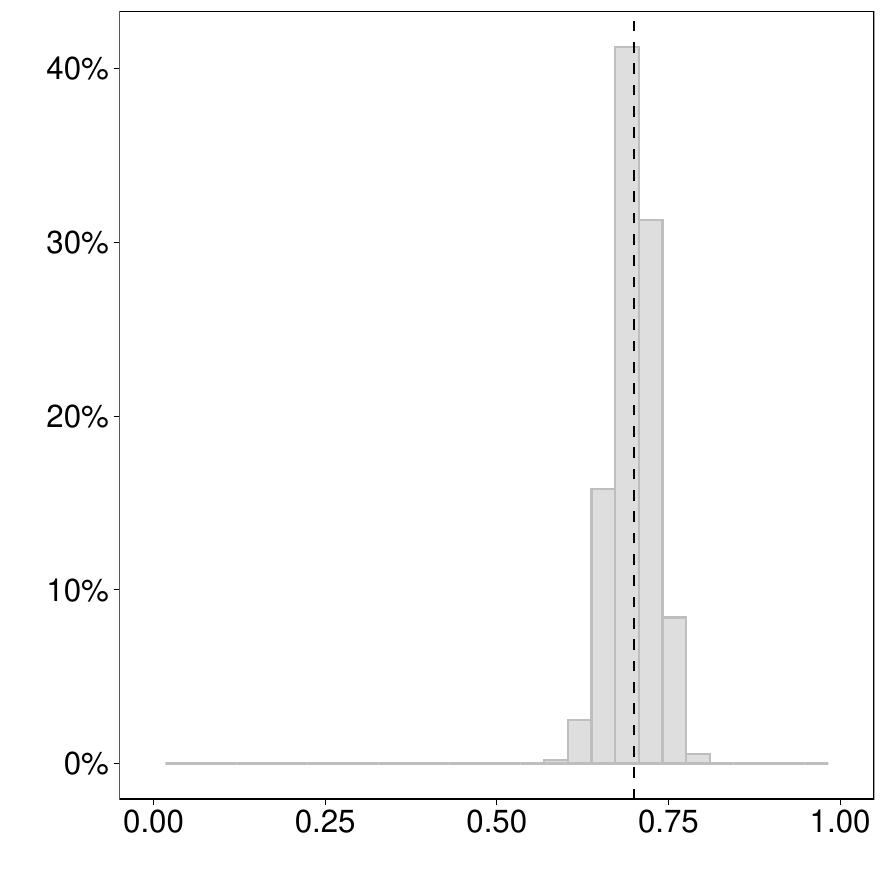} \\
FR & IT & CA \\
\includegraphics[scale=0.25]{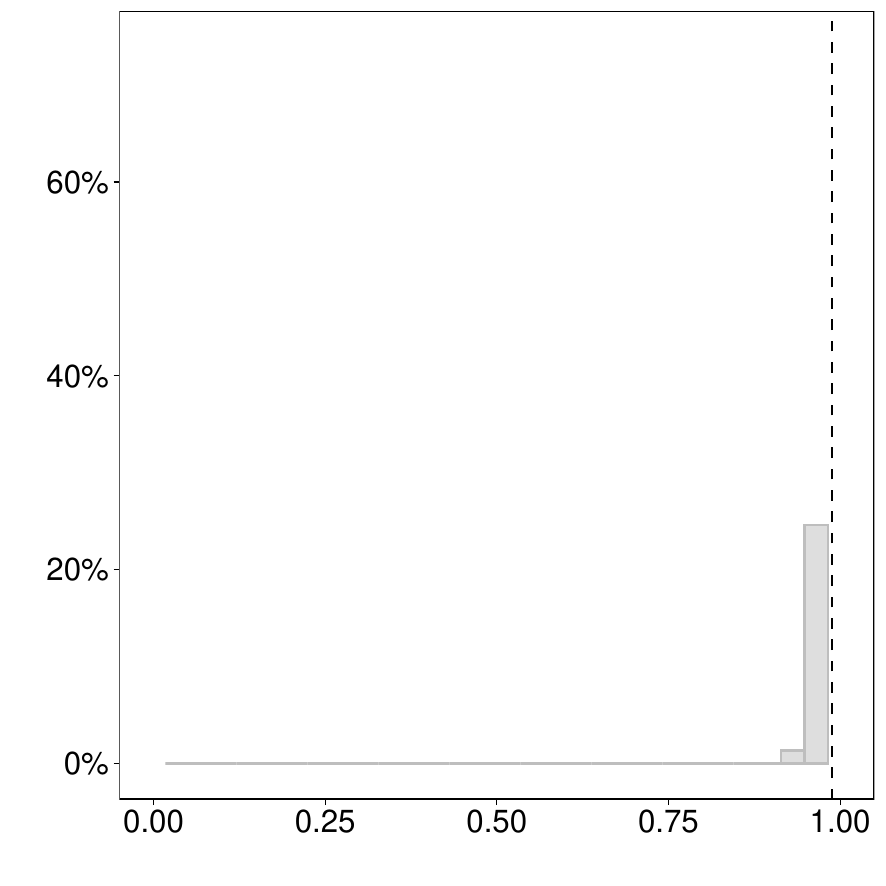} &
\includegraphics[scale=0.25]{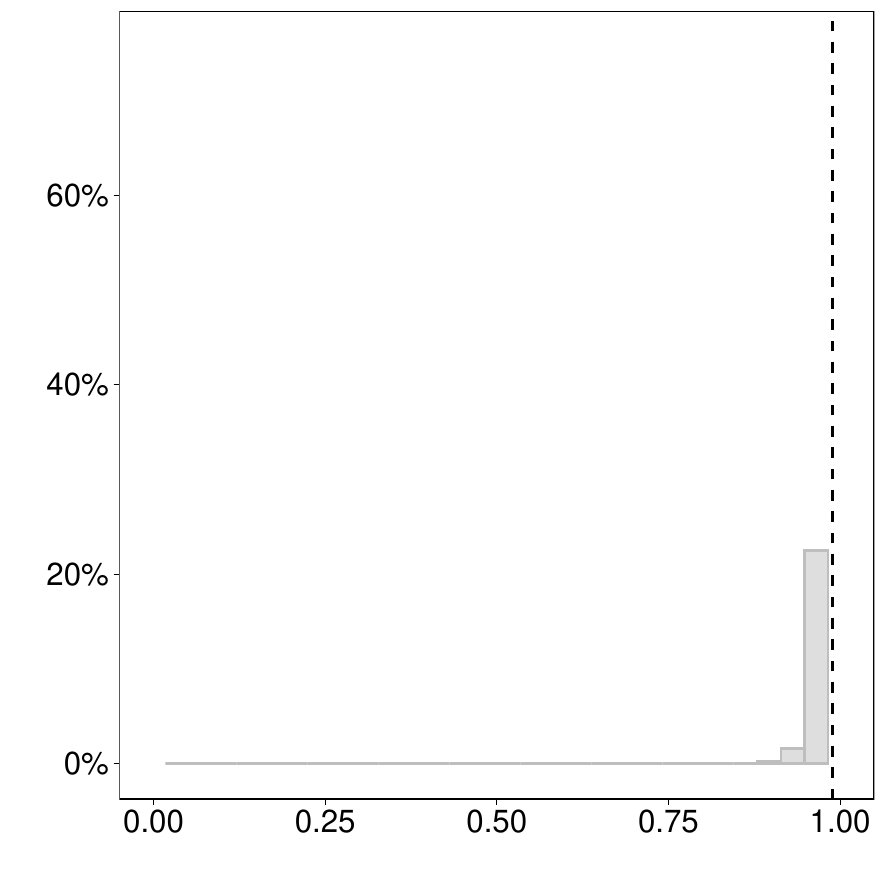} &
\includegraphics[scale=0.25]{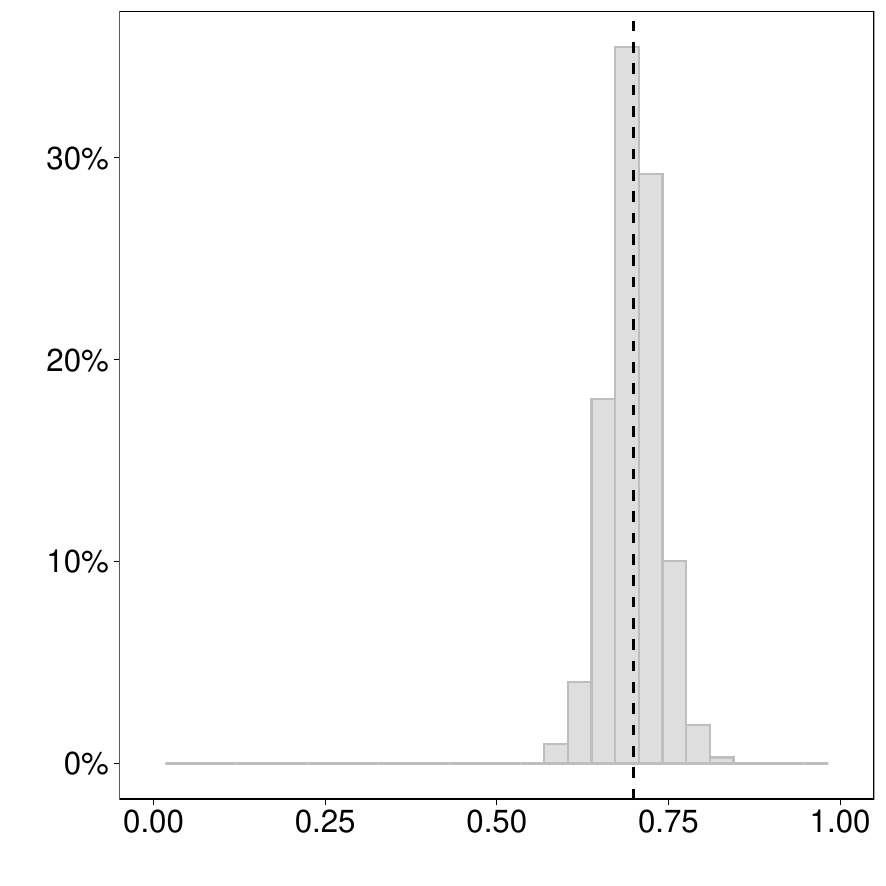} 
\end{tabular}
\caption{Posterior distribution of the country-specific spatial weight $\rho_j$, for each country $j$, with the posterior mean (dashed line) for G7 returns.}
\label{fig:insample_rho}
\end{figure}

Subsequently, we examine country-specific spatial weights, $\rho_j$, in the diagonal matrix $R$ in eq.~\eqref{eq:At_def}. The interpretation of $\rho_j$ is direct: as its value increases, so does a country's level of exposure to the network. Figure~\ref{fig:insample_rho} and Table~\ref{tab:insample_rho_ci} reveal that all exposures are positive yet display a degree of heterogeneity.
Of particular interest are the United States, which stands out as the sole G7 country displaying a non-significant exposure to the network, as indicated by the credible interval in Table~\ref{tab:insample_rho_ci}.
The seminal work of \cite{eun1989} underscores the existence of multi-lateral interactions across national stock markets and emphasises that shocks originating in the United States swiftly propagate to other stock markets. In contrast, the reverse influence is not as pronounced.
Our findings provide an additional perspective on country relationship data.
While the United States stand out as the most influential country in network centrality, as demonstrated in Figure~\ref{fig:eig}, cross-country relationships do not significantly impact its stock market.
The scenario is different for the other G7 countries. Canada, Japan, and the United Kingdom exhibit a substantial coefficient close to $0.70$, underscoring the significant exposure of their financial systems to the network. At the same time, France, Germany, and Italy show even higher levels of network exposure, with values approaching $1$. This heightened exposure may signify extensive economic interdependencies, particularly pronounced within the European Union.
OECD data indicates that the EU boasts the highest net trade in goods and services.\footnote{OECD (2023), trade in goods and services (indicator). \url{doi: 10.1787/0fe445d9-en}}
These countries share a common currency and engage in substantial cross-border trade, which can lead to a greater sensitivity to network dynamics.\footnote{The time-varying volatility, $\exp(h_{j,t})$, are included in the supplement and show as expected the heteroscedasticity commonly observed in financial data.}

\begin{table}[h!t]
\centering
\footnotesize
\begin{tabular}{c c c c c c c c} 
\toprule 
 & US & JP & DE & UK & FR & IT & CA \\
\midrule 
$\hat\rho$ & 0.1447 & \textbf{0.6891} & \textbf{0.9945} & \textbf{0.6997} & \textbf{0.9881} & \textbf{0.9885} & \textbf{0.6993} \\
CI  2.5\% & -0.0419 & 0.5235 & 0.9731 & 0.6368 & 0.9523 & 0.9505 & 0.6257 \\
CI  97.5\% & 0.2969 & 0.8452 & 1.0000 & 0.7615 & 1.0000 & 0.9999 & 0.7732 \\
\bottomrule
\end{tabular}
\caption{Posterior mean of country-specific spatial weight ($\rho_j$) for each country $j$, along with the 95\% credible interval for G7 returns. Statistically significant coefficients (those with 95\% credible intervals not including zero) are in bold.}
\label{tab:insample_rho_ci}
\end{table}

\begin{figure}[h!t]
\centering
\includegraphics[scale=0.35]{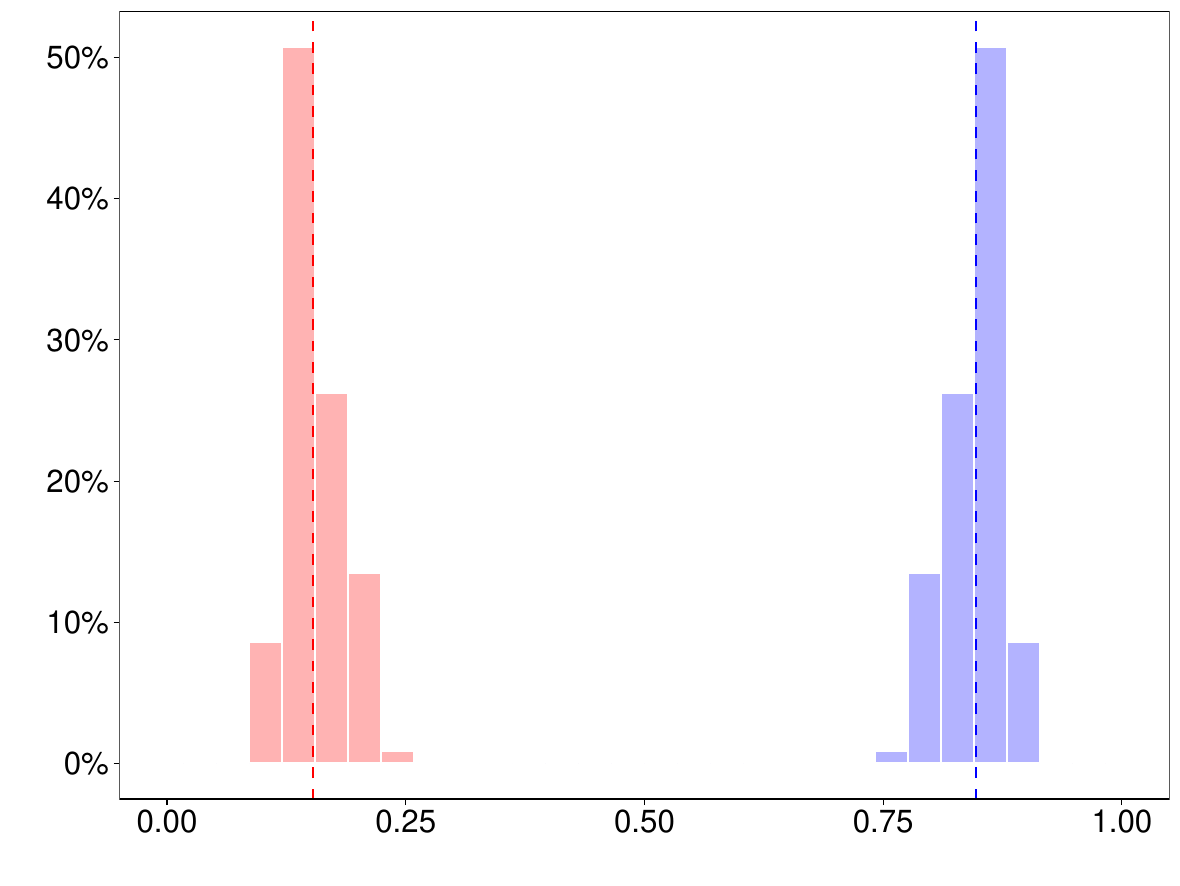}
\caption{Posterior distribution of $\bdelta$, with $\delta_1$ (weight of conflictual relationships layer) in red and $\delta_2$ (weight of cooperative relationships layer) in blue for G7 volatilities. The posterior mean is reported as a dashed line.}
\label{fig:insample_volatility_delta}
\end{figure}

\begin{figure}[h!t]
\centering
\footnotesize
\begin{tabular}{c c c}
US & JP & DE \\
\includegraphics[scale=0.25]{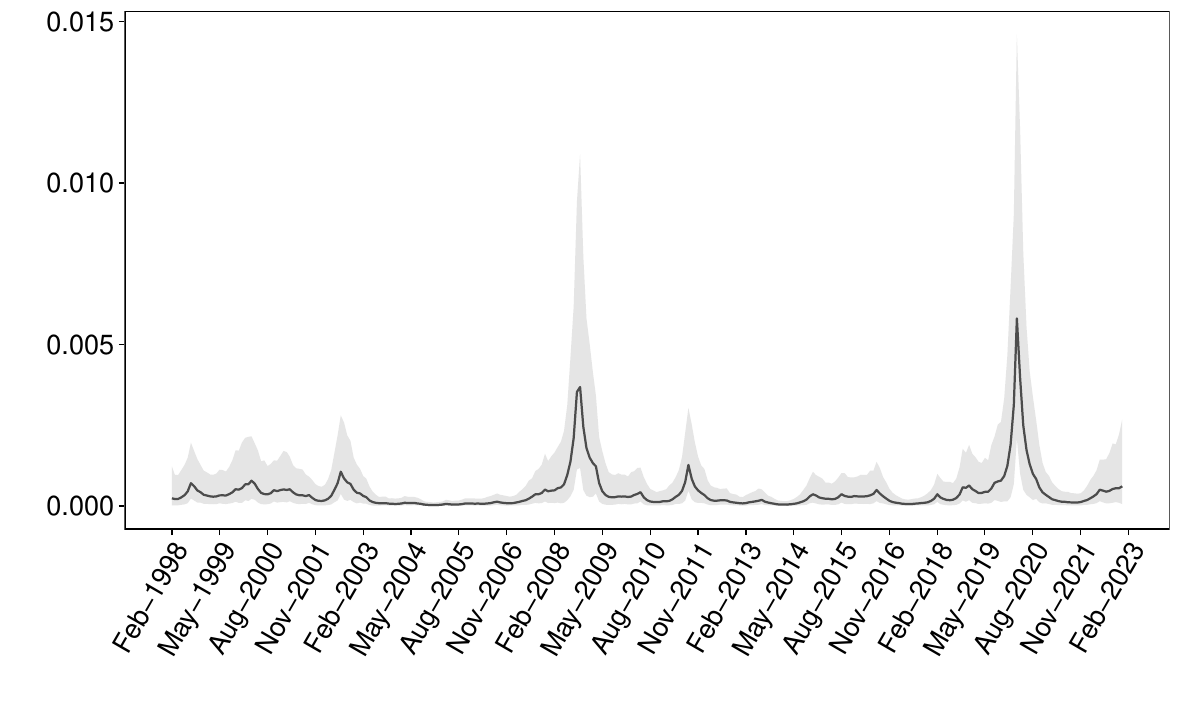} &
\includegraphics[scale=0.25]{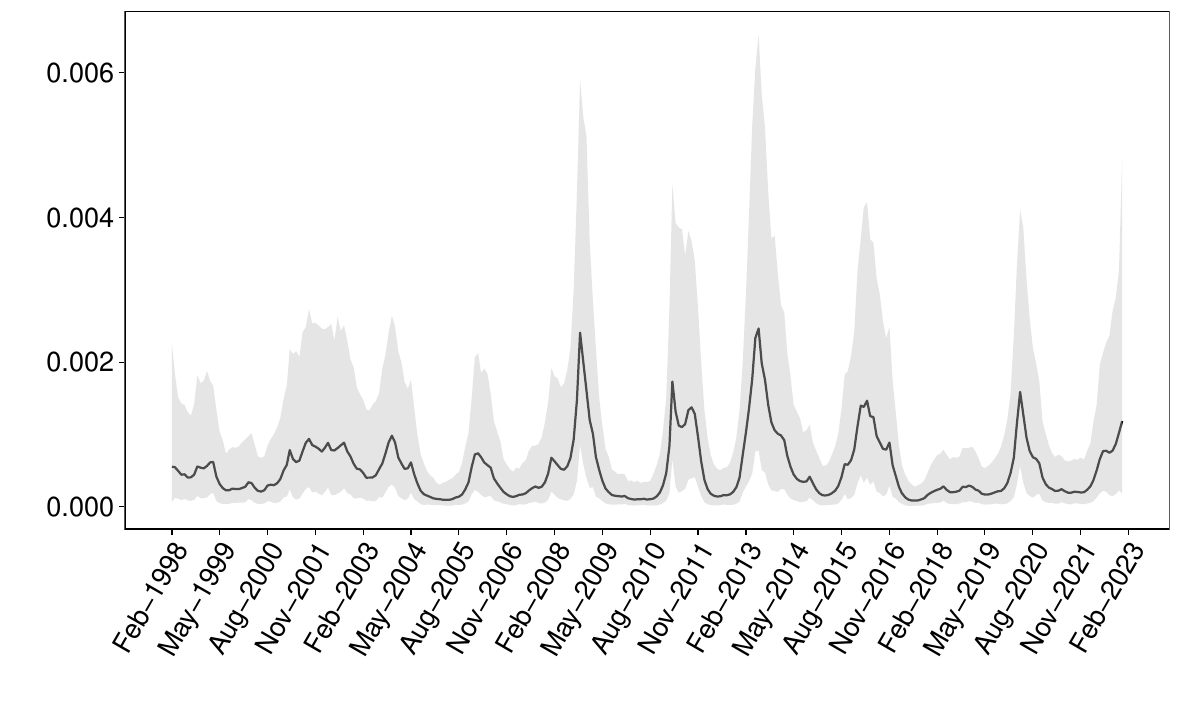} &
\includegraphics[scale=0.25]{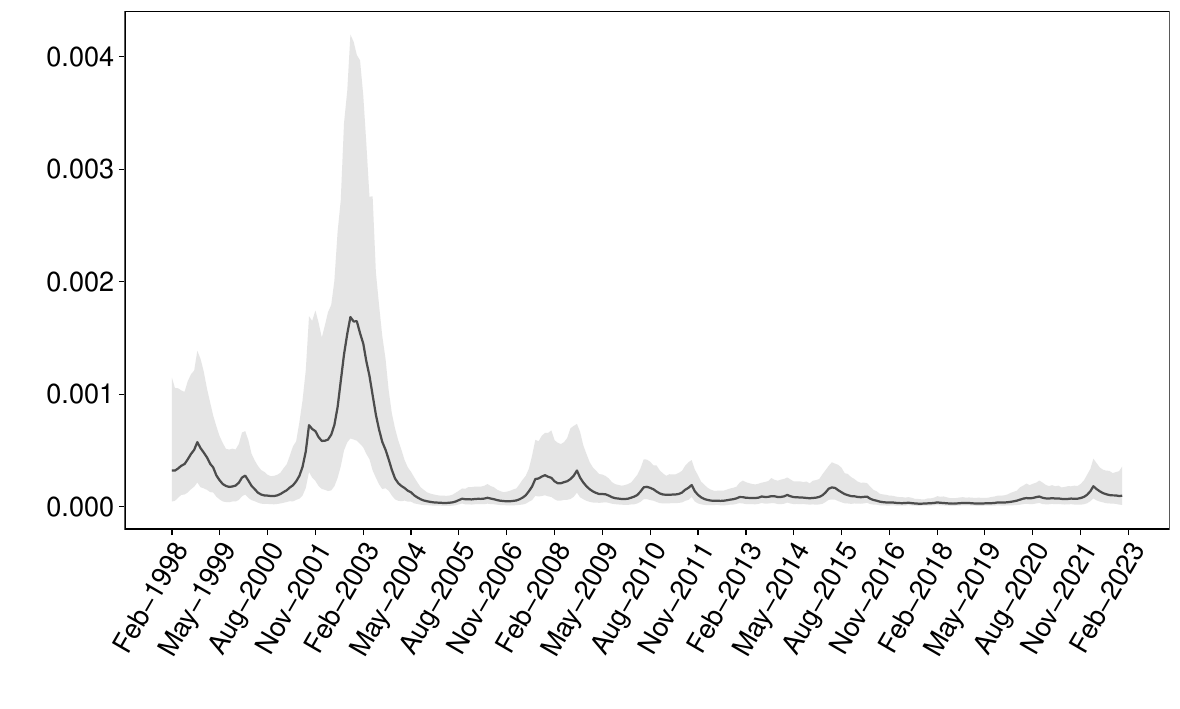} \\
UK & FR & IT \\
\includegraphics[scale=0.25]{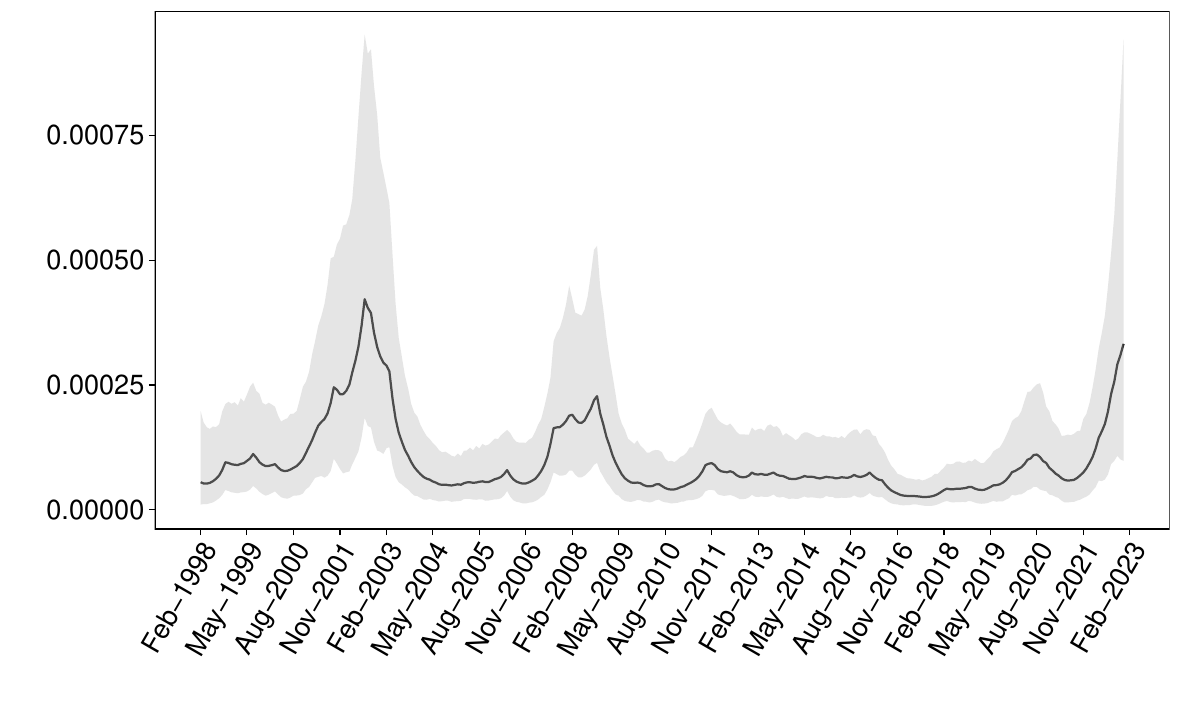} &
\includegraphics[scale=0.25]{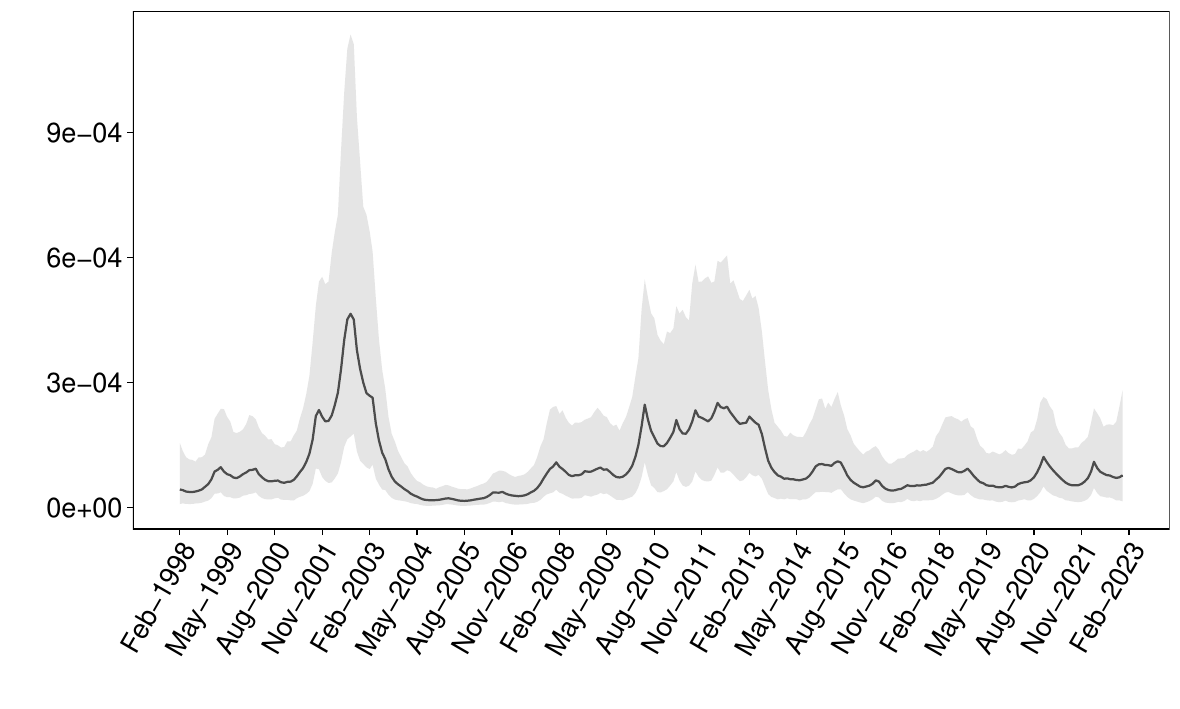} &
\includegraphics[scale=0.25]{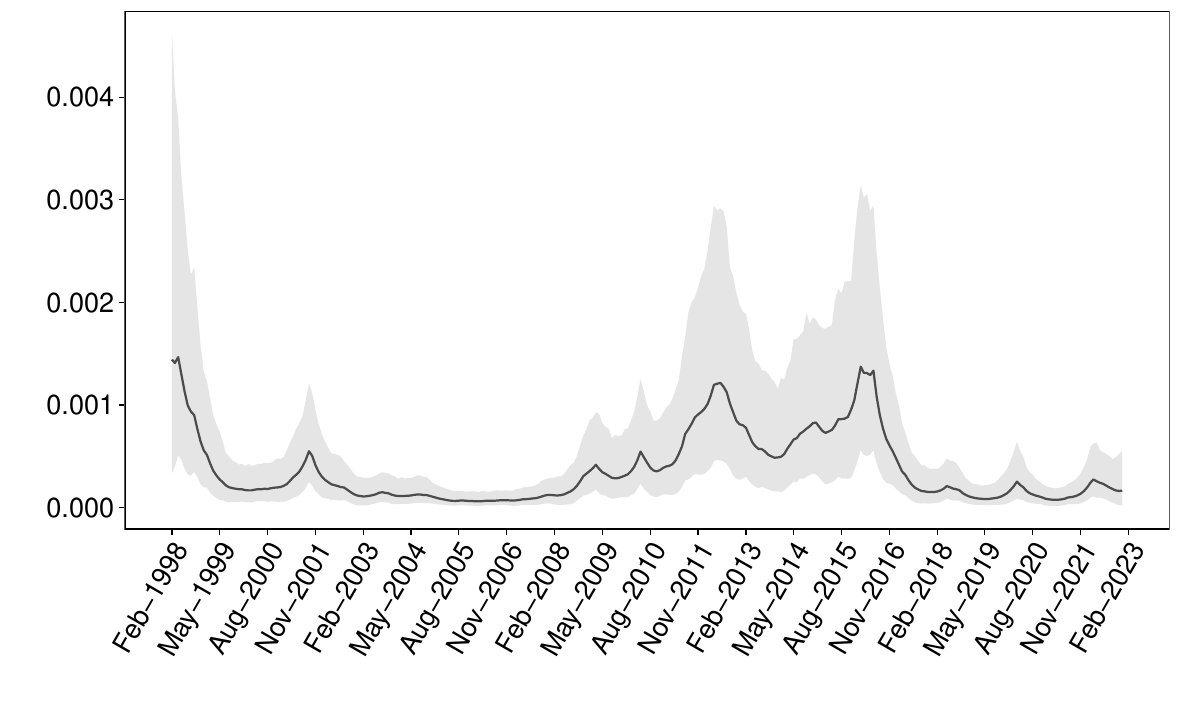} \\
CA \\
\includegraphics[scale=0.25]{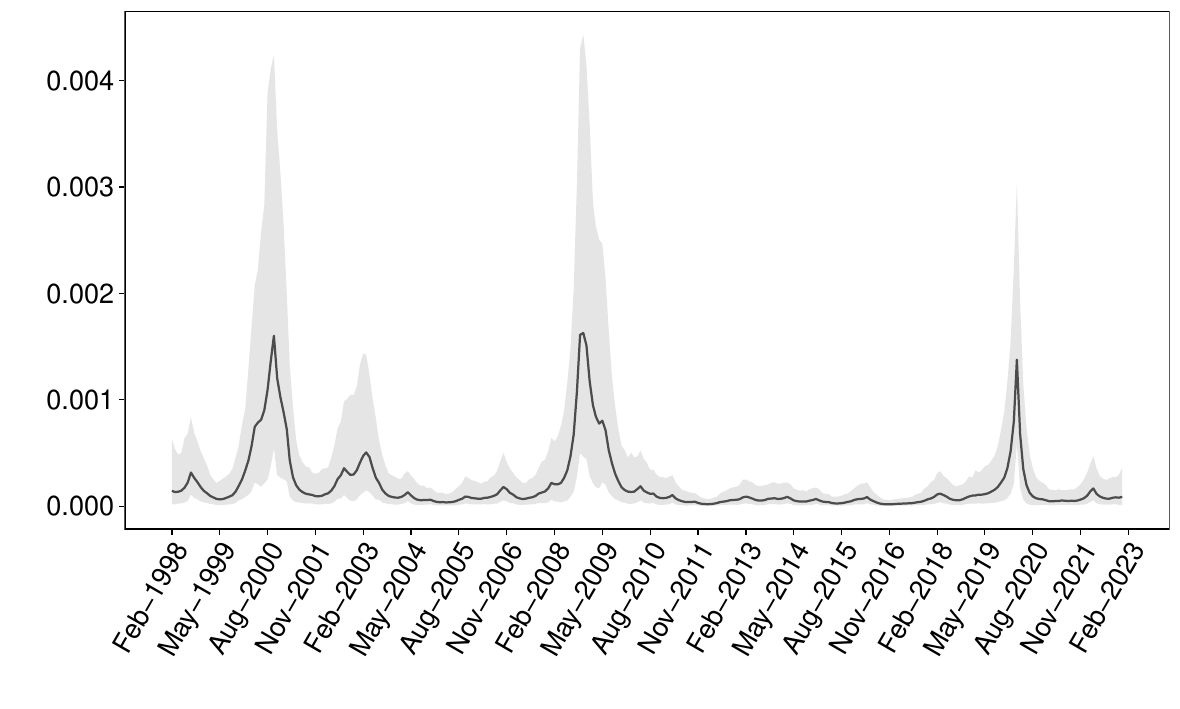} 
\end{tabular}
\caption{Posterior mean of the country-specific variance of G7 volatilities, $\exp(h_{j,t})$, for each country $j$ (line), with 95\% credible intervals (grey shade).}
\label{fig:insample_volatility_of_volatilities}
\end{figure}

Finally, we examine the G7 volatilities, where the control variable comprises the monthly realised volatility of the MSCI World Index, lagged by one month, and discuss how these results align with those of the case involving returns.
Figure \ref{fig:insample_volatility_delta} illustrates the posterior distribution of $\delta_1$ (associated with the conflictual layer) and $\delta_2$ (representing the cooperative layer) showing that the cooperative layer, with a posterior mean of $\hat\delta_2 = 0.847$, significantly outweighs the conflictual layer ($\hat\delta_1 = 0.153$) in its impact on stock market volatilities. 
As for the case of returns, this weight differential underscores the prevalence of cooperative connectedness.\footnote{The remaining results, as presented for the case of returns, are included in the Supplement.}
Table~\ref{tab:volatility_rho_ci} highlights a notable distinction from the scenario observed in returns. 
In this context, all the countries display positive and statistically significant exposures, affirming their interdependence within the G7 network. 
This suggests that while the United States is certainly influenced by the dynamics of the G7 network in terms of market risk, its level of exposure is discernibly more modest than its counterparts ($\hat\rho_1 = 0.187$).
Figure~\ref{fig:insample_volatility_of_volatilities} displays the posterior mean of each country's time-varying volatility throughout the period, showing pronounced spikes during specific country-related events.  
This aligns with the findings drawn by \cite{grassi2015s}, who demonstrated that the dynamic behaviour of realised volatility series during financial turbulence is associated with structural changes in the parameters guiding the stochastic volatility model.

\begin{table}[h!t]
\centering
\footnotesize
\begin{tabular}{c c c c c c c c} 
\toprule 
 & US & JP & DE & UK & FR & IT & CA \\
\midrule 
$\hat\rho$ & \textbf{0.1869} & \textbf{0.6546} & \textbf{0.8827} & \textbf{0.7434} & \textbf{0.9498} & \textbf{0.9316} & \textbf{0.6487} \\
CI  2.5\% & 0.0770 & 0.5427 & 0.8373 & 0.6949 & 0.9062 & 0.8412 & 0.5887 \\
CI  97.5\% & 0.2794 & 0.7916 & 0.9239 & 0.7919 & 0.9927 & 0.9984 & 0.7094 \\
\bottomrule
\end{tabular}
\caption{Posterior mean of country-specific spatial weight ($\rho_j$) for each country $j$, along with the 95\% credible interval for G7 volatilities. Statistically significant coefficients (those with 95\% credible intervals not including zero) are in bold.}
\label{tab:volatility_rho_ci}
\end{table}

\subsubsection{G7 spillover effects}
\begin{figure}[h!]
\centering
\captionsetup{width=0.92\linewidth}
\includegraphics[scale=0.40]{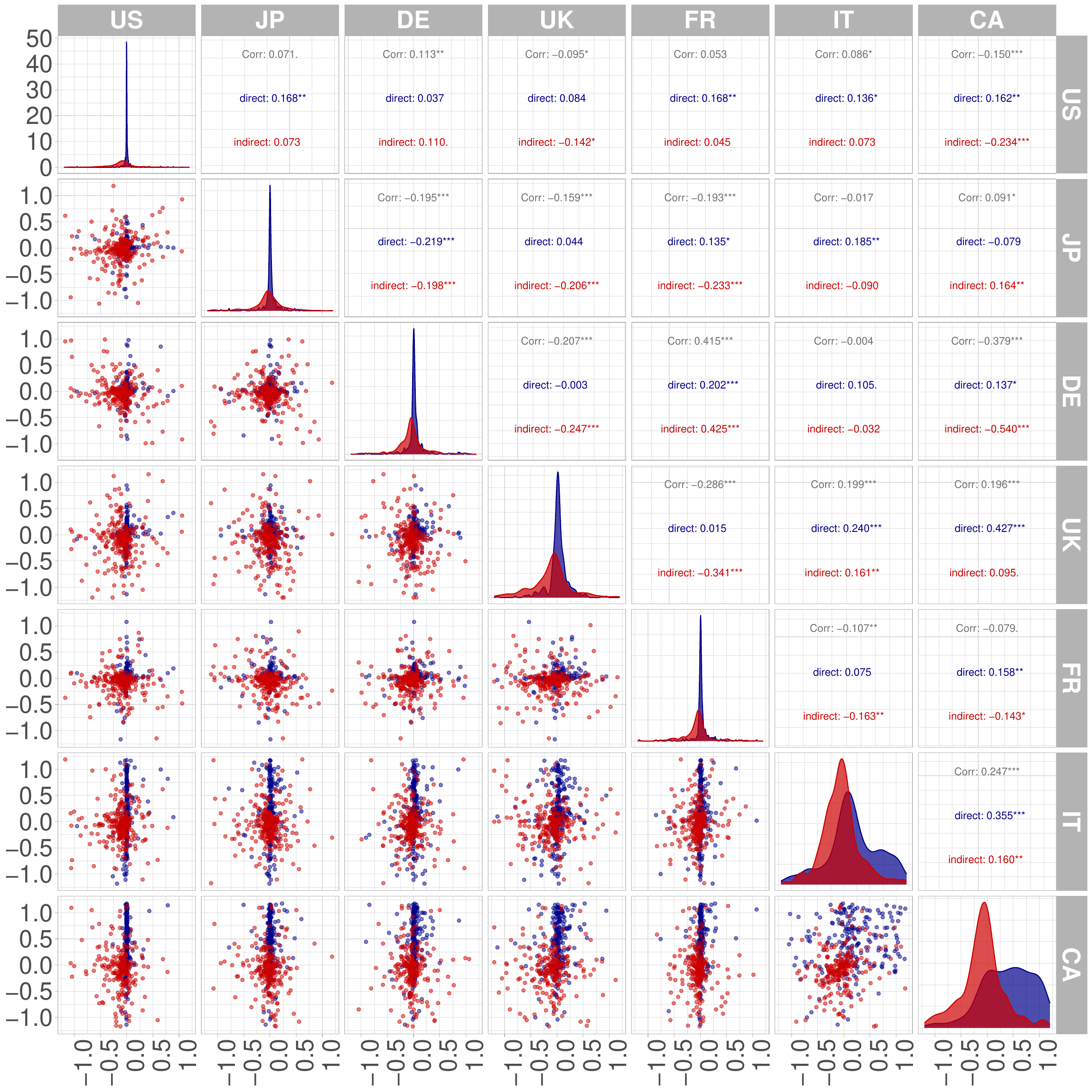} 
\caption{Upper: pairwise cross-correlation of overall (grey), of direct (blue) and indirect (red) spillover effects over time between any pair of countries. Stars denote significance at $10\%$ ($^{*}$), $5\%$ ($^{**}$), $1\%$ ($^{***}$).
Diagonal: kernel density estimate of the distribution of direct (blue) and indirect (red) effects over time for each country, visualised on the interval $(-1.2,1.2)$.
Bottom: scatter of the bivariate distribution of direct (blue) and indirect (red) effects over time for each pair of countries, visualised on the region $(-1.2,1.2) \times (-1.2,1.2)$.}
\label{fig:insample_crosscorr_effects_returns}
\end{figure}

In the following, we consider the spillover effects as outlined in eq.~\eqref{eq:direct_effects} and \eqref{eq:indirect_effects}. 
In Figure~\ref{fig:insample_crosscorr_effects_returns}, the diagonal displays kernel density estimates of the distribution of direct (blue) and indirect (red) spillover effects over time for each country in the case of returns. These are visually represented within the interval $(-1.2, 1.2)$.\footnote{The case of volatilities presents analogous results and are included in the Supplement.}
Interestingly, Canada, followed by Italy and the United Kingdom, exhibit the highest spillover variability over time for both the direct and the indirect effect.\footnote{In the supplement, we also include the box plots of the distribution over time of the country-specific spillover effects.}
Canada is heavily dependent on commodities in terms of exports, and this variability could be ascribed to energy commodity shocks that occurred in the last decades, such as the oil price plunge of 2014-2016, the COVID-19 pandemic, and the Russian invasion of Ukraine in 2022.
The empirical findings by \cite{reboredo2021} support this observation. Their research underscores that a diverse array of commodities, spanning agriculture, energy, industrial metals, livestock, and precious metals, are pivotal conduits for extensive spillovers between these commodities and the Canadian stock market.
Italy is the sole member of the G7 group to have weathered a sovereign debt crisis in 2011. This crisis left a profound imprint on its banking sector, escalating systemic risk across European financial markets \citep{black2016systemic}.
Turning our attention to the United Kingdom, many factors may have played a role. Foremost among these was the seismic event of Brexit in 2016, which ushered in heightened political uncertainty and substantially impacted financial markets \citep{hudson2020}.

In the upper panels, cross-correlations between overall (grey), direct (blue), and indirect (red) effects over time for all country pairs are depicted. 
These indices provide insight into the interdependence structure of spillovers among various G7 countries, revealing notable distinctions when dissecting the direct and indirect effects. For instance, in the case of the United States, all countries, except Germany and the United Kingdom, exhibit an overall positive and statistically significant correlation. This trend is also mirrored in the direct spillover effects. However, when it comes to the indirect effect, a notably significant negative correlation emerges, particularly with Canada.
Conversely, Germany showcases a negative overall significant correlation with the United Kingdom and Italy. However, a positive correlation with the latter becomes apparent upon closer examination of the direct effect, while the indirect effect yields a negative correlation.
This phenomenon aligns with findings in \cite{ehrmann2017euro}, which identified a negative yield correlation between Italy and Germany in the period spanning from March 2010 to March 2012. 
Similarly, there is a negative correlation for the indirect spillover effects in the case of Italy and France.
Overall, the analysed dependencies within the G7 nations highlight the intricate dynamics between direct and indirect spillover effects, which often exhibit opposite signs.
This phenomenon can be attributed to various key factors, including heterogeneous market expectations, policy responses, and underlying risk factors.


Finally, we delve into the relationship between spillover effects and the Economic Policy Uncertainty (EPU) index proposed by \cite{baker2016}. 
As noted by the authors, policy uncertainty increases during financial crises and policy disputes. 
The aim is to verify if countries' spillovers in the G7 networks are related to policy uncertainty.
The correlations between spillover effects and their respective EPU indices for the G7 countries are summarised in Table \ref{tab:EPU_corr_level_ret}.
While examining returns, we find that none of the countries demonstrate a significant correlation with the EPU index. However, when it comes to volatilities, a compelling observation emerges. The United States exhibits a substantial negative correlation (-0.1531) with a significance level of 1\%. This indicates that during periods of heightened economic uncertainty, the impact of the G7 network on the United States is notably reduced.
This finding is particularly noteworthy as it aligns with previous results emphasising the influential role of the United States within the G7 group (refer to Figure \ref{fig:eig}). 
Additionally, it corroborates the earlier observation of lower but still significant exposure to the G7 network for the United States (see Table \ref{tab:volatility_rho_ci}).

\begin{table}[h!t]
\centering
\footnotesize
\begin{tabular}{c c c c c c c c} 
\toprule 
 & US & JP & DE & UK & FR & IT & CA \\
 \midrule 
returns &-0.0291&0.0343&0.0178&-0.0045&0.0583&-0.0260&-0.0070\\
volatilities &-0.1531$^{***}$&0.0058&-0.0236&0.0531&-0.0087&0.0384&0.0122\\
\bottomrule
\end{tabular}
\caption{Correlation between the spillover effects of each country and their respective EPU index proposed by \cite{baker2016} for returns and volatilities. Stars denote significance at $10\%$ ($^{*}$), $5\%$ ($^{**}$), $1\%$ ($^{***}$).}
\label{tab:EPU_corr_level_ret}
\end{table}

\FloatBarrier
\section{Conclusions} \label{sec:conclusion}

This study introduced a novel spatial autoregressive model tailored for panel data, which incorporates time-varying networks within a multilayer structure and country-specific network exposures.
Our approach also allows for the temporal variation of the structural variance and enables the analysis of the impact of shocks in terms of both direct and indirect spillover effects.
By leveraging the spatial multiplier matrix, our approach measures the impact of shocks in terms of both direct and indirect spillover effects.

Empirically, we focused on the G7 economies, examining the dynamics of their network relationships and their impact on stock markets. 
Our findings reveal that cooperative interactions substantially influence stock market returns more than conflictual interactions within the G7. Moreover, country-specific spatial weights highlight varying levels of exposure to the network, with implications for systemic risk and financial stability. 
The spillover analysis underscores the complex interplay between direct and indirect spillover effects, highlighting critical factors influencing market dynamics.

Future research avenues may extend this framework to other observed economic and financial networks and delve deeper into specific network dynamics in countries' relationships.
For instance, it would be interesting to generalise our approach to dynamic spatial panel models \citep[e.g., see][]{kuersteiner2020dynamic}.

\bibliographystyle{apalike}
\bibliography{biblio}

\begin{thebibliography}{}

\bibitem[Abiad and Qureshi, 2023]{abiad2023}
Abiad, A. and Qureshi, I.~A. (2023).
\newblock The macroeconomic effects of oil price uncertainty.
\newblock {\em Energy Economics}, 125:106839.

\bibitem[Ahlgren and Antell, 2017]{ahlgren2017tests}
Ahlgren, N. and Antell, J. (2017).
\newblock Tests for abnormal returns in the presence of event-induced
  cross-sectional correlation.
\newblock {\em Journal of Financial Econometrics}, 15(2):286--301.

\bibitem[Anselin, 1988]{anselin1988spatial}
Anselin, L. (1988).
\newblock {\em Spatial econometrics: Methods and models}, volume~4.
\newblock Springer Science \& Business Media.

\bibitem[Anselin et~al., 2008]{anselin2008spatial}
Anselin, L., Gallo, J.~L., and Jayet, H. (2008).
\newblock Spatial panel econometrics.
\newblock In M\'{a}ty\'{a}s, L. and Sevestre, P., editors, {\em The
  econometrics of panel data: Fundamentals and recent developments in theory
  and practice}, pages 625--660. Springer.

\bibitem[Baker et~al., 2016]{baker2016}
Baker, S.~R., Bloom, N., and Davis, S.~J. (2016).
\newblock Measuring economic policy uncertainty.
\newblock {\em The Quarterly Journal of Economics}, 131(4):1593--1636.

\bibitem[Baker et~al., 2020]{baker2020}
Baker, S.~R., Bloom, N., Davis, S.~J., Kost, K., Sammon, M., and Viratyosin, T.
  (2020).
\newblock The unprecedented stock market reaction to covid-19.
\newblock {\em The Review of Asset Pricing Studies}, 10(4):742--758.

\bibitem[Bhadra et~al., 2016]{bhadra2016default}
Bhadra, A., Datta, J., Polson, N.~G., and Willard, B. (2016).
\newblock Default {Bayesian} analysis with global-local shrinkage priors.
\newblock {\em Biometrika}, 103(4):955--969.

\bibitem[Bhattacharya et~al., 2015]{bhattacharya2015dirichlet}
Bhattacharya, A., Pati, D., Pillai, N.~S., and Dunson, D.~B. (2015).
\newblock {Dirichlet}--{Laplace} priors for optimal shrinkage.
\newblock {\em Journal of the American Statistical Association},
  110(512):1479--1490.

\bibitem[Billio et~al., 2023]{billio2023impact}
Billio, M., Caporin, M., Panzica, R., and Pelizzon, L. (2023).
\newblock The impact of network connectivity on factor exposures, asset
  pricing, and portfolio diversification.
\newblock {\em International Review of Economics \& Finance}, 84:196--223.

\bibitem[Billio et~al., 2012]{billio2012econometric}
Billio, M., Getmansky, M., Lo, A.~W., and Pelizzon, L. (2012).
\newblock Econometric measures of connectedness and systemic risk in the
  finance and insurance sectors.
\newblock {\em Journal of Financial Economics}, 104(3):535--559.

\bibitem[Black et~al., 2016]{black2016systemic}
Black, L., Correa, R., Huang, X., and Zhou, H. (2016).
\newblock The systemic risk of european banks during the financial and
  sovereign debt crises.
\newblock {\em Journal of Banking \& Finance}, 63:107--125.

\bibitem[Bonaccolto et~al., 2019]{bonaccolto2019estimation}
Bonaccolto, G., Caporin, M., and Panzica, R. (2019).
\newblock Estimation and model-based combination of causality networks among
  large us banks and insurance companies.
\newblock {\em Journal of Empirical Finance}, 54:1--21.

\bibitem[Caldara and Iacoviello, 2022]{caldara2022}
Caldara, D. and Iacoviello, M. (2022).
\newblock Measuring geopolitical risk.
\newblock {\em American Economic Review}, 112(4):1194--1225.

\bibitem[Carvalho et~al., 2010]{carvalho2010horseshoe}
Carvalho, C.~M., Polson, N.~G., and Scott, J.~G. (2010).
\newblock The horseshoe estimator for sparse signals.
\newblock {\em Biometrika}, 97(2):465--480.

\bibitem[Corrado and Fingleton, 2012]{corrado2012economics}
Corrado, L. and Fingleton, B. (2012).
\newblock Where is the economics in spatial econometrics?
\newblock {\em Journal of Regional Science}, 52(2):210--239.

\bibitem[Debarsy and LeSage, 2018]{debarsy2018flexible}
Debarsy, N. and LeSage, J. (2018).
\newblock Flexible dependence modeling using convex combinations of different
  types of connectivity structures.
\newblock {\em Regional Science and Urban Economics}, 69:48--68.

\bibitem[Debarsy and LeSage, 2022]{debarsy2022bayesian}
Debarsy, N. and LeSage, J.~P. (2022).
\newblock Bayesian model averaging for spatial autoregressive models based on
  convex combinations of different types of connectivity matrices.
\newblock {\em Journal of Business \& Economic Statistics}, 40(2):547--558.

\bibitem[Diebold and Y{\i}lmaz, 2014]{diebold2014network}
Diebold, F.~X. and Y{\i}lmaz, K. (2014).
\newblock On the network topology of variance decompositions: Measuring the
  connectedness of financial firms.
\newblock {\em Journal of Econometrics}, 182(1):119--134.

\bibitem[Dittrich et~al., 2017]{dittrich2017bayesian}
Dittrich, D., Leenders, R. T.~A., and Mulder, J. (2017).
\newblock Bayesian estimation of the network autocorrelation model.
\newblock {\em Social Networks}, 48:213--236.

\bibitem[Ehrmann and Fratzscher, 2017]{ehrmann2017euro}
Ehrmann, M. and Fratzscher, M. (2017).
\newblock Euro area government bonds--fragmentation and contagion during the
  sovereign debt crisis.
\newblock {\em Journal of International Money and Finance}, 70:26--44.

\bibitem[Elhorst, 2003]{elhorst2003specification}
Elhorst, J.~P. (2003).
\newblock Specification and estimation of spatial panel data models.
\newblock {\em International Regional Science Review}, 26(3):244--268.

\bibitem[Engle et~al., 2020]{engle2020}
Engle, R.~F., Giglio, S., Kelly, B., Lee, H., and Stroebel, J. (2020).
\newblock Hedging climate change news.
\newblock {\em The Review of Financial Studies}, 33(3):1184--1216.

\bibitem[Eun and Shim, 1989]{eun1989}
Eun, C.~S. and Shim, S. (1989).
\newblock International transmission of stock market movements.
\newblock {\em Journal of Financial and Quantitative Analysis}, 24(2):241--256.

\bibitem[Gerner et~al., 2002]{gerner2002conflict}
Gerner, D.~J., Schrodt, P.~A., Yilmaz, O., and Abu-Jabr, R. (2002).
\newblock Conflict and mediation event observations ({CAMEO}): {A} new event
  data framework for the analysis of foreign policy interactions.
\newblock {\em International Studies Association, New Orleans}.

\bibitem[Grassi and de~Magistris, 2015]{grassi2015s}
Grassi, S. and de~Magistris, P.~S. (2015).
\newblock It's all about volatility of volatility: Evidence from a two-factor
  stochastic volatility model.
\newblock {\em Journal of Empirical Finance}, 30:62--78.

\bibitem[Hu et~al., 2023]{hu2023arbitrage}
Hu, J., Ding, H., and Liu, X. (2023).
\newblock Arbitrage pricing with heterogeneous spatial effects and
  heteroscedastic disturbances.
\newblock {\em Journal of Financial Econometrics}, 21(4):1169--1195.

\bibitem[Hudson et~al., 2020]{hudson2020}
Hudson, R., Urquhart, A., and Zhang, H. (2020).
\newblock Political uncertainty and sentiment: Evidence from the impact of
  brexit on financial markets.
\newblock {\em European Economic Review}, 129:103523.

\bibitem[Jackson and Pernoud, 2021]{jackson2021systemic}
Jackson, M.~O. and Pernoud, A. (2021).
\newblock Systemic risk in financial networks: A survey.
\newblock {\em Annual Review of Economics}, 13:171--202.

\bibitem[Kastner and Fr{\"u}hwirth-Schnatter, 2014]{kastner2014ancillarity}
Kastner, G. and Fr{\"u}hwirth-Schnatter, S. (2014).
\newblock Ancillarity-sufficiency interweaving strategy ({ASIS}) for boosting
  {MCMC} estimation of stochastic volatility models.
\newblock {\em Computational Statistics \& Data Analysis}, 76:408--423.

\bibitem[Kelejian and Piras, 2016]{kelejian2016extension}
Kelejian, H.~H. and Piras, G. (2016).
\newblock An extension of the {J}-test to a spatial panel data framework.
\newblock {\em Journal of Applied Econometrics}, 31(2):387--402.

\bibitem[Kuersteiner and Prucha, 2020]{kuersteiner2020dynamic}
Kuersteiner, G.~M. and Prucha, I.~R. (2020).
\newblock Dynamic spatial panel models: {Networks}, common shocks, and
  sequential exogeneity.
\newblock {\em Econometrica}, 88(5):2109--2146.

\bibitem[Lee and Liu, 2010]{lee2010efficient}
Lee, L.-f. and Liu, X. (2010).
\newblock Efficient gmm estimation of high order spatial autoregressive models
  with autoregressive disturbances.
\newblock {\em Econometric Theory}, 26(1):187--230.

\bibitem[LeSage and Pace, 2009]{lesage2009introduction}
LeSage and Pace (2009).
\newblock {\em Introduction to spatial econometrics}.
\newblock Chapman and Hall/CRC.

\bibitem[LeSage and Parent, 2007]{lesage2007bayesian}
LeSage and Parent (2007).
\newblock Bayesian model averaging for spatial econometric models.
\newblock {\em Geographical Analysis}, 39(3):241--267.

\bibitem[Neal, 2003]{neal2003slice}
Neal, R.~M. (2003).
\newblock Slice sampling.
\newblock {\em The Annals of Statistics}, 31(3):705--767.

\bibitem[Ord, 1975]{ord1975estimation}
Ord, K. (1975).
\newblock Estimation methods for models of spatial interaction.
\newblock {\em Journal of the American Statistical Association},
  70(349):120--126.

\bibitem[Polson and Scott, 2010]{polson2010shrink}
Polson, N.~G. and Scott, J.~G. (2010).
\newblock Shrink globally, act locally: {Sparse} {Bayesian} regularization and
  prediction.
\newblock {\em Bayesian Statistics}, 9(501-538):105.

\bibitem[Reboredo et~al., 2021]{reboredo2021}
Reboredo, J.~C., Ugolini, A., and Hernandez, J.~A. (2021).
\newblock Dynamic spillovers and network structure among commodity, currency,
  and stock markets.
\newblock {\em Resources Policy}, 74:102266.

\bibitem[Sun et~al., 1999]{sun1999posterior}
Sun, D., Tsutakawa, R.~K., and Speckman, P.~L. (1999).
\newblock Posterior distribution of hierarchical models using {CAR}(1)
  distributions.
\newblock {\em Biometrika}, 86(2):341--350.

\bibitem[Taylor, 1994]{taylor1994modeling}
Taylor, S.~J. (1994).
\newblock Modeling stochastic volatility: A review and comparative study.
\newblock {\em Mathematical Finance}, 4(2):183--204.

\bibitem[Yang and Lee, 2021]{yang2021estimation}
Yang, K. and Lee, L.-f. (2021).
\newblock Estimation of dynamic panel spatial vector autoregression:
  {Stability} and spatial multivariate cointegration.
\newblock {\em Journal of Econometrics}, 221(2):337--367.

\bibitem[Yu et~al., 2012]{yu2012estimation}
Yu, J., de~Jong, R., and Lee, L.-f. (2012).
\newblock Estimation for spatial dynamic panel data with fixed effects: {The}
  case of spatial cointegration.
\newblock {\em Journal of Econometrics}, 167(1):16--37.

\bibitem[Zhang and Yu, 2018]{zhang2018spatial}
Zhang, X. and Yu, J. (2018).
\newblock Spatial weights matrix selection and model averaging for spatial
  autoregressive models.
\newblock {\em Journal of Econometrics}, 203(1):1--18.

\end{thebibliography}

\end{document}